\documentclass[12pt]{myelsart}
\usepackage{epsfig}
\usepackage{cite,mcite}

\def\GeV{\ifmmode {\mathrm{\ Ge\kern -0.1em V}}\else
                   \textrm{Ge\kern -0.1em V}\fi}%
\def\MeV{\ifmmode {\mathrm{\ Me\kern -0.1em V}}\else
                   \textrm{Me\kern -0.1em V}\fi}%
\def\um{\ifmmode  {\mathrm{\mu m}}\else \textrm{$\mu$m}\fi}%

\def\D0{D\O{}}

\begin{document}
\pagestyle{plain}
\pagenumbering{arabic}
\setcounter{page}{1}

%============================================================Title page
\begin{frontmatter}
% \begin{titlepage}
\date{12 July 2000}

\title{Cosmic ray tests of the \D0 preshower detector}

  \author[Kansas]    {P. Baringer},
  \author[FNAL]      {A. Bross},
  \author[Rochester] {V. Buescher},
  \author[Rochester] {F. Canelli},
  \author[Rochester] {G. Davis},
  \author[Umich]     {K. Del Signore},
  \author[Stony]     {S. Desai},
  \author[Rochester] {J. Estrada},
  \author[Rochester] {G. Ginther},
  \author[BNL]       {A. Gordeev},
  \author[Stony]     {P. Grannis},
  \author[FNAL]      {S. Gr\"unendahl},
  \author[Umich]     {S. Hou\thanksref{corresp}},
  \author[BNL]       {J. Kotcher},
  \author[FNAL]      {D. Lincoln},
  \author[BNL]       {M. Liu},
  \author[Indiana]   {A.A. Mayorov},
  \author[Umich]     {H.A. Neal},
  \author[FNAL]      {T. Nunnemann},
  \author[Stony]     {A. Patwa},
  \author[Umich]     {J. Qian},
  \author[Stony]     {M. Rijssenbeek},
  \author[FNAL]      {P. Rubinov},
  \author[LTU]       {L. Sawyer},
  \author[Stony]     {A. Talalaevskii},
  \author[Umich]     {A.S. Turcot},
  \author[Indiana]   {R. Van Kooten},
  \author[Stony]     {Z.M. Wang},
  \author[NotreDame] {J. Warchol},
  \author[NotreDame] {M. Wayne},
  \author[BNL]       {P. Yamin},
  \author[FNAL]      {K. Yip},
  \author[Umich]     {B. Zhou}

  \address[Kansas]   {University of Kansas,
                      Lawrence, Kansas 66045 }
  \address[FNAL]     {Fermi National Accelerator Laboratory,
                      Batavia, Illinois 60510}
  \address[Rochester]{University of Rochester, Rochester,
                      New York 14627}
  \address[Umich]    {University of Michigan, Ann Arbor,
                      Michigan 48109}
  \address[Stony]    {State University of New York,
                      Stony Brook, New York 11794}
  \address[BNL]      {Brookhaven National Laboratory, Upton,
                      New York 11973}
  \address[Indiana]  {Indiana University, Bloomington,
                      Indiana 47405}
  \address[LTU]      {Louisiana Tech University,
                      Ruston, Louisiana 71272}
  \address[NotreDame]{University of Notre Dame, Notre Dame,
                      Indiana 46556}

  \thanks[corresp]{e-mail: suen@fnal.gov}

  {\large   For the \D0 Collaboration }

\begin{abstract}

The \D0 preshower detector consists of scintillator strips 
with embedded wavelength-shifting fibers,
and a readout using Visible Light Photon Counters.
The response to minimum ionizing particles has been tested
with cosmic ray muons.
We report results on the gain calibration and light-yield distributions.
The spatial resolution is investigated taking into account the light
sharing between strips, the effects of multiple scattering and
various systematic uncertainties.
The detection efficiency and noise contamination are also investigated.

\end{abstract}

\vspace{.5cm} 
\end{frontmatter}
% \end{titlepage}

%================================================Section 1 Introduction
\section{Introduction}

The \D0 Central and Forward Preshower (CPS and FPS) detectors 
\cite{D0} employ scintillator strips with embedded 
wavelength-shifting (WLS) fibers and readout via Visible Light Photon 
Counters (VLPCs) \cite{VLPC}.
The CPS has three concentric cylindrical 
layers of strips covering a 
pseudorapidity\addtocounter{footnote}{-1}\footnotemark\
region of $|\eta|<1.3$.
It is located outside a solenoid coil and lead absorber, 
which comprise a total of two radiation lengths ($2 X_0$) at
$\eta$=0.
The FPS consists of trapezoidally-shaped modules in front and 
behind a $2X_0$ lead absorber, covering the pseudorapidity region
$1.5 < |\eta| <2.5$.
The preshower detectors sample the energy deposition with
fine granularity, and thereby provide 
information for the identification of electrons and photons.

\footnotetext{
  Pseudorapidity is defined as $\eta = - \ln \tan (\theta /2)$,
  where $\theta$ is the polar angle with respect to the proton beam.
\vspace{-1cm} }

\begin{figure}[h!]                     %----------------------Fig.1
  \centering
  \includegraphics[width=.72\linewidth] {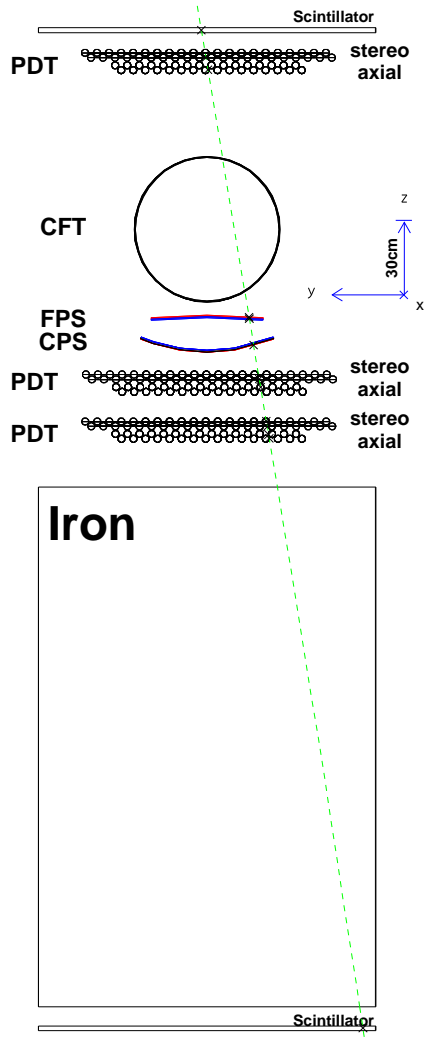}  %{bench.eps}
  \vspace{-.3cm}
  \caption{ An end view of the cosmic ray test facility is
            illustrated by an event display from a GEANT simulation.
  \label{fig:bench} }
\end{figure}                           %----------------------Fig.1

Studies of the prototype detectors comparing different construction 
techniques have been previously reported \cite{sqstrip,tristrip}.
In this paper, we report cosmic ray tests of a representative 
axial module of the CPS and of a FPS module, which were studied
at Fermilab during three periods from August 1999 to March 2000.
We first describe the apparatus and the data processing,
the gain calibration of photoelectrons 
and the light yield for minimum ionizing particles (MIPs).
The spatial resolution is investigated by examining
light sharing between neighboring strips.
The systematic uncertainties and effects from multiple scattering 
are estimated using GEANT 3 simulations \cite{GEANT}.
Finally, the detection efficiency and noise are examined
as a function of readout threshold.

\begin{figure}[b!]                     %----------------------Fig.2
  \centering
  \includegraphics[width=.52\linewidth] {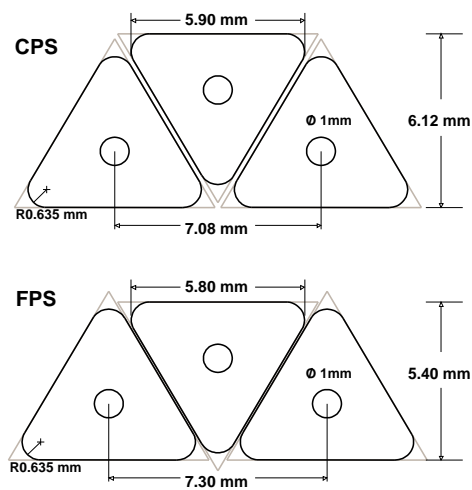}
  \vspace{-1.cm}
  \caption{Cross sections of CPS and FPS scintillator strips.
           A wavelength-shifting fiber is embedded in the strip
           center where the circle is drawn.
  \label{fig:strip} }
  \vspace{-.8cm}
\end{figure}                           %----------------------Fig.2

\begin{figure}[t!]                     %----------------------Fig.3
  \vspace{-1.2cm}
  \centering
  \includegraphics[width=.52\linewidth] {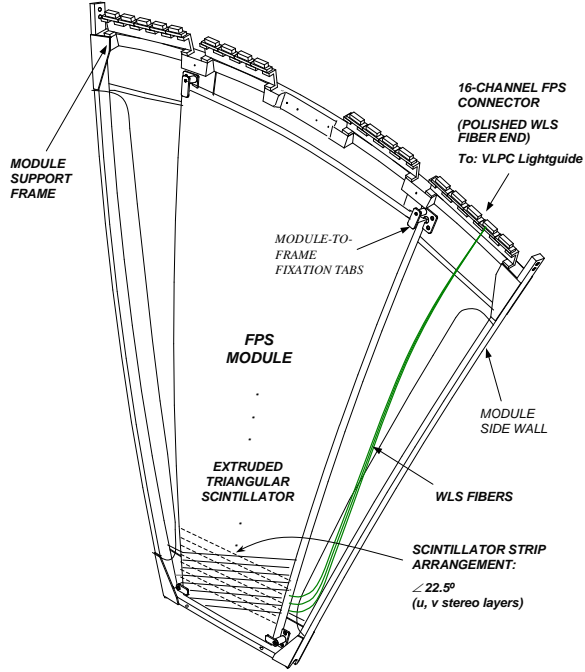}
  \vspace{-.5cm}
  \caption{Schematic view of a FPS module, which consists of
           two layers of scintillator strips.
  \label{fig:fps} }
  \vspace{.2cm}
\end{figure}                           %----------------------Fig.3

%=============================================================Section 2
\section {Apparatus}

The experimental setup is illustrated in Fig.~\ref{fig:bench}.
Cosmic ray muons are filtered by iron blocks having
a 2.5 \GeV{} equivalent stopping power, and triggered
by the coincidence of scintillation counters above
and below the apparatus.
The reference tracking system consists of three sets of proportional 
drift tubes (PDTs) \cite{PDT}: each with an axial
and a stereo layer of 32 tubes crossing at a $6^\circ$ angle.
The trigger area is covered fully by the PDTs, which have 25 mm pitch 
and are 2800 mm long.
The test detectors include an axial CPS module, and a FPS module.
A prototype of the third cylinder of 
the \D0 central fiber tracker (CFT) \cite{D0} is also present.
Its performance is reported elsewhere.

The preshower scintillator layers consist of interlocking
triangular strips, as illustrated in Fig.~\ref{fig:strip}.
The strips of equilateral triangular cross section 
are made by extrusion of polystyrene plastic \cite{extrusion},
which is doped with 1\% p-terphenyl and 150 ppm diphenyl stilbene,
similar to the Bicron \cite{Bicron} BC-404 scintillator.
Each scintillator strip is wrapped in aluminized mylar 
for optical isolation.  
The packing density is different for the CPS and the FPS modules,
which results in different layer thickness and strip pitches.
The strip pitch is 3.54 mm for the axial CPS module 
and 3.65 mm for the FPS.

Embedded at the center of the scintillator strip is a WLS fiber 
which collects and directs the light to the edge of the module,
where a clear light-guide fiber transports light
to the input of the VLPC.
The WLS fibers are Kuraray \cite{Kuraray}
Y-11 (250 ppm concentration) multi-clad fibers of 835 \um{} diameter.
The non-readout ends are silvered for better light collection.
The clear fibers are Kuraray multi-clad S-type fibers 
of 835 \um{} diameter, and the lengths used in this test
are 11~m and 13.5~m.

The CPS module is an octant of a cylinder of average radius 721.5 mm
containing one layer of 160 scintillator strips that are
2412 mm long.
The WLS fibers of half the strip length are inserted from 
both ends of the strips, meet in the middle,
and thus they divide the module into 
the {\it north} and the {\it south} segments for readout.
Each segment has the middle 64 strips connected to the readout system.
The scintillator strips are aligned with the axial PDT tubes.
The strip positions at the module edges differ in vertical ($z$)
direction by 55 mm from the strips at the center of the module.

The FPS module (illustrated in Fig.~\ref{fig:fps})
has two scintillator layers (denoted by $u$ and $v$);
each has 144 strips of lengths between 104 mm to 293 mm.
The upper 128 strips are connected to the readout.
The strips of the FPS-$u$ layer are aligned to the axial PDT
and the strips of the FPS-$v$ make a 22.5$^\circ$ angle with 
the $u$-layer strips.
The scintillator layers have a curvature of average radius
2540 mm, and the strip positions in $z$ vary 
by a maximum of 12 mm.

\begin{figure}[b!]                     %----------------------Fig.4
  \vspace{-.5cm}
  \centering
  \includegraphics[width=.70\linewidth] {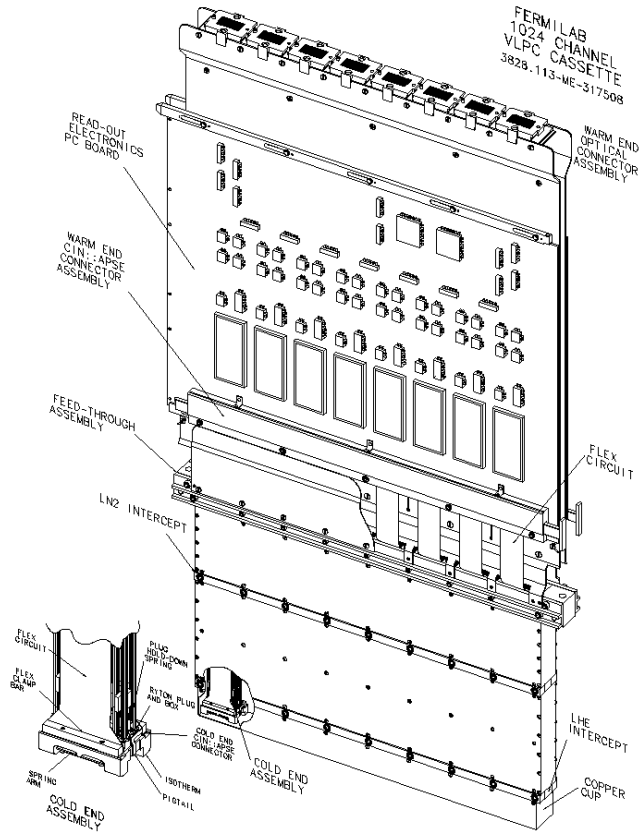}
  \vspace{-.5cm}
  \caption{ VLPC cassette assembly.
  \label{fig:VLPC} }
  \vspace{-1.cm}
\end{figure}                           %----------------------Fig.4

\begin{figure}[t!]                     %----------------------Fig.5
  \vspace{-.8cm}
  \centering
  \includegraphics[width=.75\linewidth] {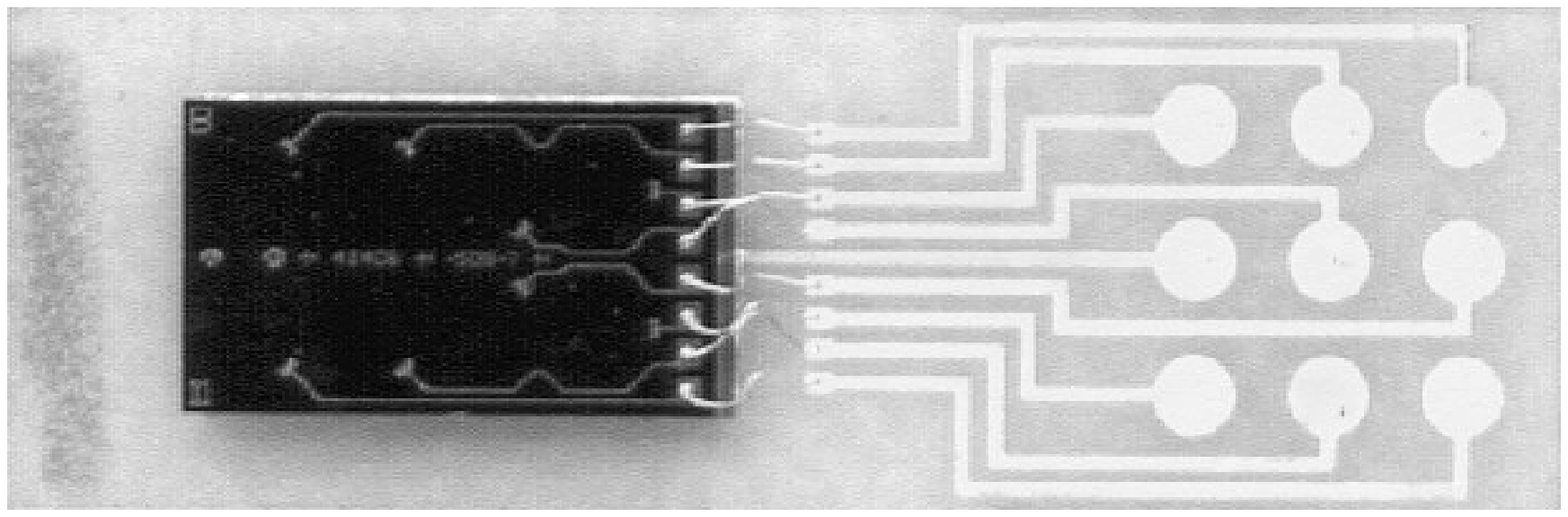}
  \vspace{-1.cm}
  \caption{ Picture of a VLPC hybrid.
            The VLPC chip has 8 pixels of 1 mm diameter.
            The gold-plate pads on the substrate are the contacts to
            the flex cable for the 8 signal outputs and the bias.
  \label{fig:chip} }
  \vspace{.5cm}
% \end{figure}                           %----------------------Fig.5
% \begin{figure}[!h]                     %----------------------Fig.6
  \vspace{-.5cm}
  \centering
  \includegraphics[width=.65\linewidth] {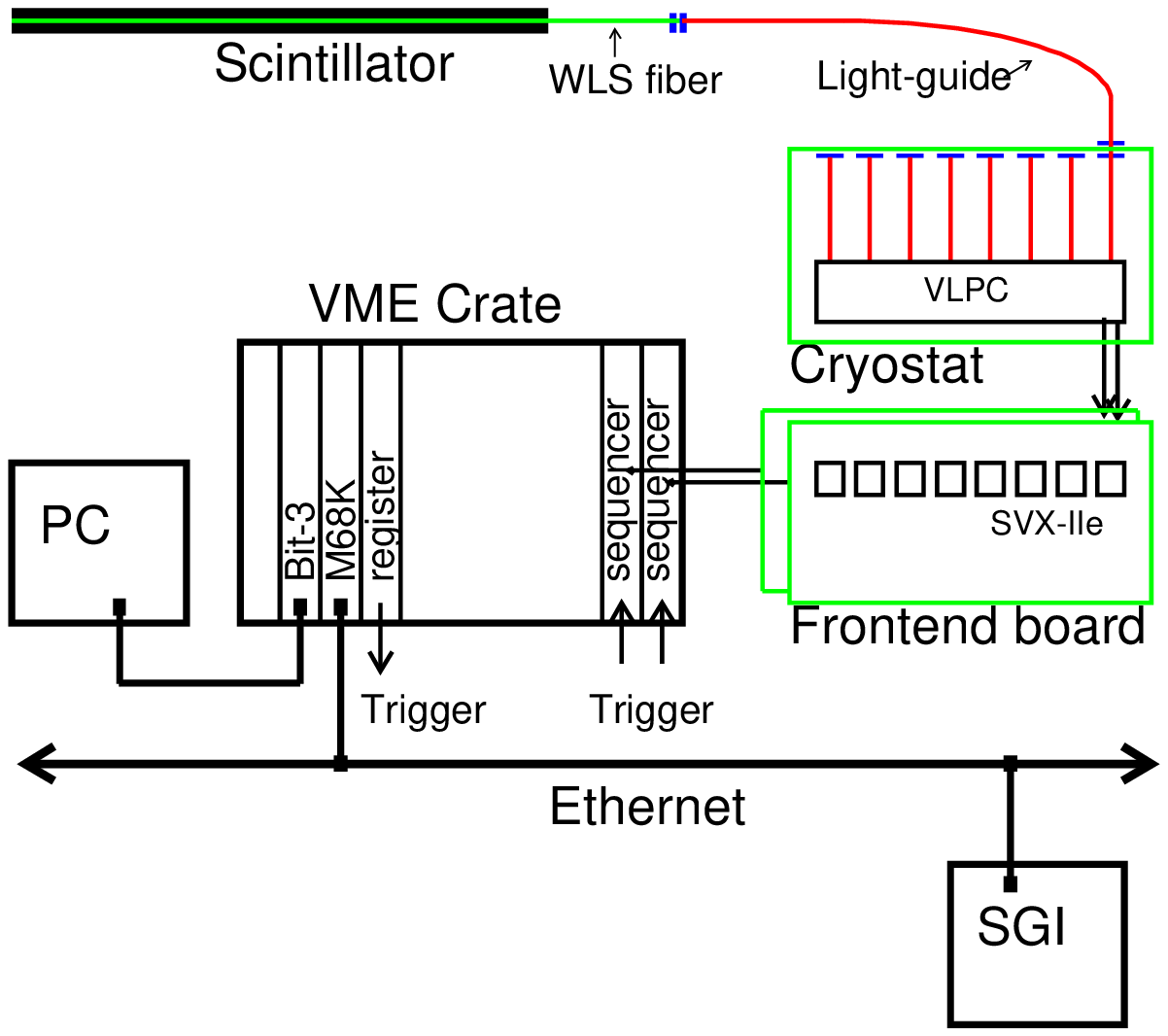}
  \vspace{-2.5cm}
  \caption{ Flow diagram for the light collection and the 
            VLPC readout.
  \label{fig:lab3daq} }
  \vspace{.8cm}
\end{figure}                           %----------------------Fig.6

The signals are processed through VLPC cassettes located
in a cryostat container maintained at 9 Kelvin \cite{cryo}.
VLPCs are solid state photodetectors with a high quantum
efficiency and gain.
The quantum efficiency is bias-voltage dependent and is reported 
by the manufacturer to plateau at 80\% at WLS wavelengths.
Likewise, the gain is voltage dependent.
At the operating voltage used in this paper, the gain ranges
from 25000 to 50000.

The VLPC assembly and the readout electronics are illustrated 
in Fig.~\ref{fig:VLPC}.
A VLPC cassette has eight readout modules, each of which 
is attached to one bundle of clear fibers containing 128 fibers.
A VLPC module consists of two columns of VLPC chips,
each chip has 8 pixels (shown in Fig.~\ref{fig:chip}),
and the column of 8 chips (64 pixels)
is connected via a flex cable to a packaged SVX-IIe \cite{SVX}
chip mounted on a front-end board.
Two front-end boards are employed, each having 8 SVX chips.

The SVX chip has 128 channels, of which 64 channels are employed.
The signal from the VLPC is processed through a preamplifier,
an analog pipeline, a pedestal comparator, and a Wilkinson
8-bit ADC.
The digitized signals are sent to a stand-alone 
sequencer \cite{sequencer} in a VME crate.
The SVX chip is configured for readout of all channels
and the preamplifier dynamic range is set for 120.2 fC (0.47 fC/ADC).
The data acquisition system consists of a VME 68k CPU board
and a SGI workstation running a DART\cite{DART} based program.
The readout electronics and the data flow are
illustrated in Fig.~\ref{fig:lab3daq}.

%=============================================================Section 3
\section {Data processing}

%=====================================
\subsection {SVX pedestal}

\begin{figure}[t!]                      %----------------------Fig.7
  \vspace{-1.cm}
  \centering
  \includegraphics[width=.55\linewidth] {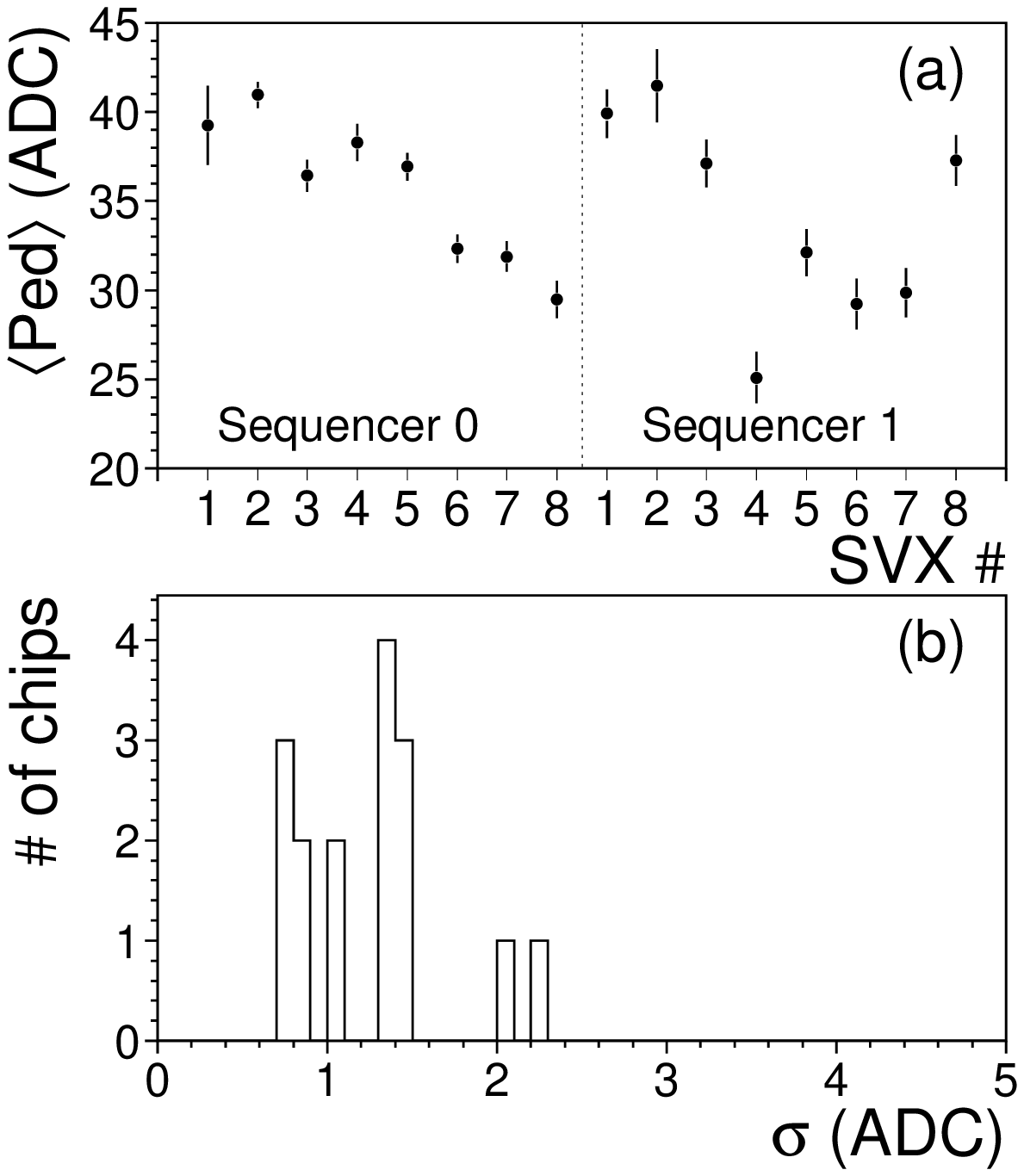}
  \vspace{-.5cm}
  \caption{
    (a) The mean pedestal for 64 channels of each SVX chips,
    and (b) the distribution of standard deviations (error bars in (a))
    indicating the significance of common-mode shift.
  \label{fig:pedcn} }
  \vspace{0.5cm}
% \end{figure}                            %----------------------Fig.7
% \begin{figure}[t!]                      %----------------------Fig.8
  \vspace{-1.cm}
  \centering
  \includegraphics[width=.55\linewidth] {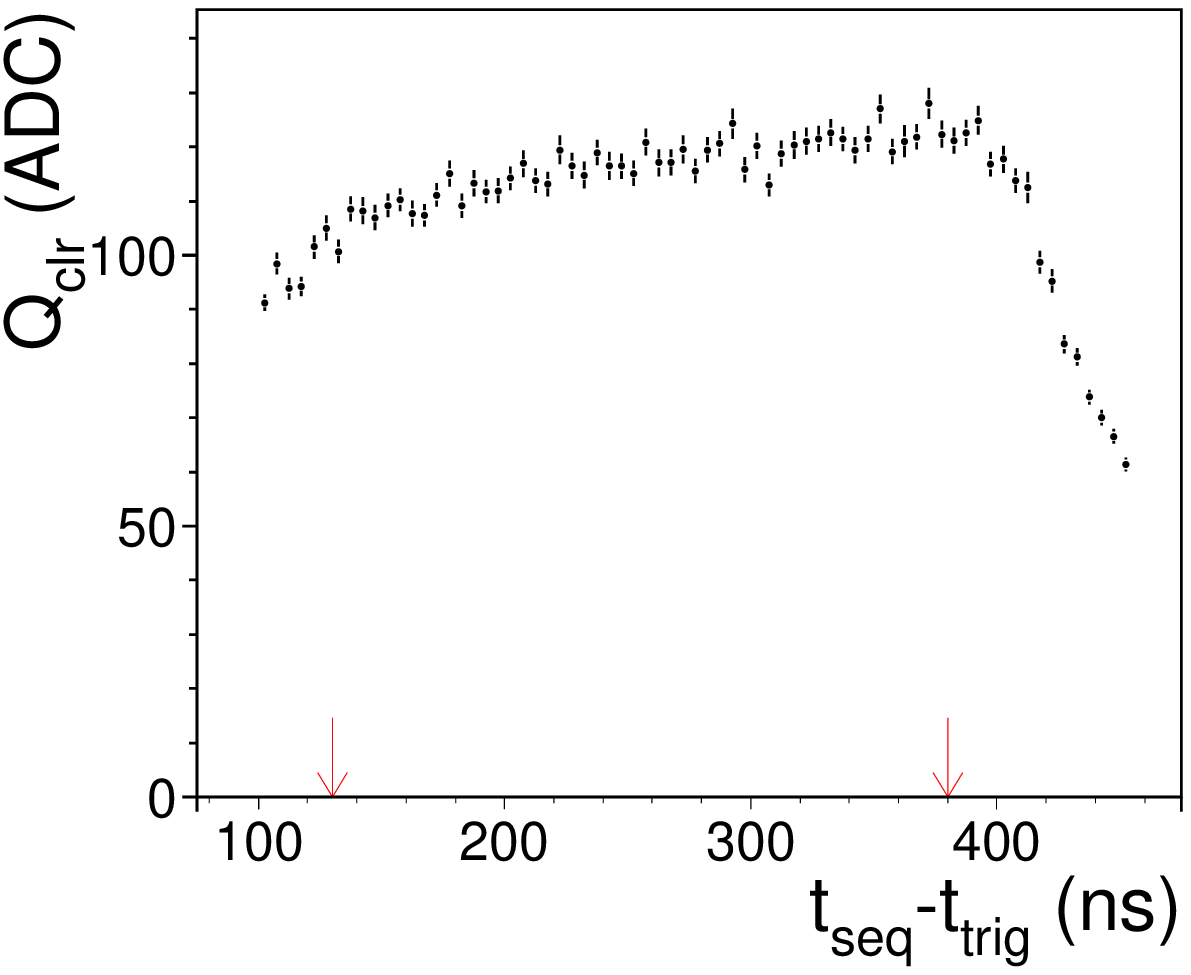}
  \vspace{-.5cm}
  \caption{ Charge on a pipeline capacitor as a function of the 
            difference in time between the cosmic ray trigger 
            and the beginning of the SVX clock cycle.
  \label{fig:qt0} }
  \vspace{.5cm}
\end{figure}

The SVX data are corrected for pedestal values and 
for common-mode shifts.
Within the same SVX chip, the pedestal values of the 64 readout channels
are quite uniform, with the standard deviation from the mean 
being typically a few ADC counts.
The common-mode shift is the pickup from low frequency noise
that shifts the pedestal.
It is calculated for each event excluding channels that may
contain a signal ($3\sigma$ above pedestal).
The mean values of the pedestals for all the 16 chips are plotted 
in Fig.~\ref{fig:pedcn}.a.
The standard deviation on the mean pedestal, which reflects
the significance of the common-mode shift, is presented
in Fig.~\ref{fig:pedcn}.b.
It is observed to be on the order of one to three ADC counts.

                            %----------------------Fig.8

%=====================================
\subsection {Timing of cosmic ray trigger}

The SVX-IIe circuits perform sequential sampling of the preamplifier
output, and feed one of the 32 storage capacitors in 
the analog pipeline.
The clock cycles are produced by a stand-alone sequencer at 
396 ns intervals, and the cosmic ray triggers are randomly 
distributed within the clock cycle.
The difference in time between the event trigger and the 
beginning of the SVX data acquisition cycle is recorded by a TDC module.
The signal contained within a pipeline capacitor
(shown in Fig.~\ref{fig:qt0}) has a relatively flat plateau
extending over 250 ns.
Event selection requires the TDC stamp to be within 
the interval between 130 to 380 ns.

\begin{figure}[b!]                       %----------------------Fig.9
  \centering
  \vspace{-.5cm}
  \includegraphics[width=.65\linewidth] {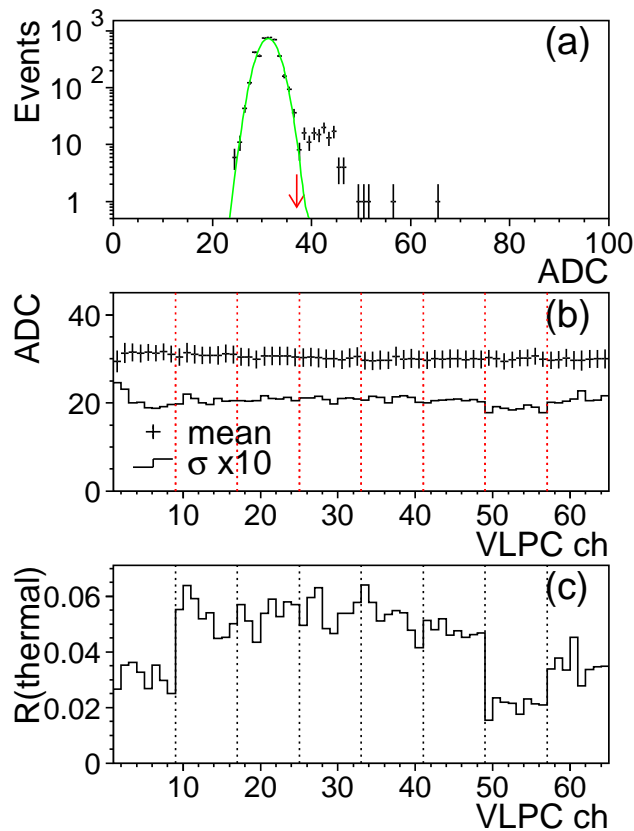}
  \vspace{-.8cm}
  \caption{ (a) Output spectrum from a VLPC channel,
            with a Gaussian fit to the pedestal peak.
            (b) The mean and the standard deviation of the Gaussian
            fits for the 64 channels of one SVX chip, and (c)
            the fraction of thermal electrons
            (events at $3\sigma$ above the mean of pedestal).
  \label{fig:led1g} }
  \vspace{-.5cm}
\end{figure}                             %----------------------Fig.9

\begin{figure}[b]                       %----------------------Fig.10
  \centering
  \includegraphics[width=.65\linewidth] {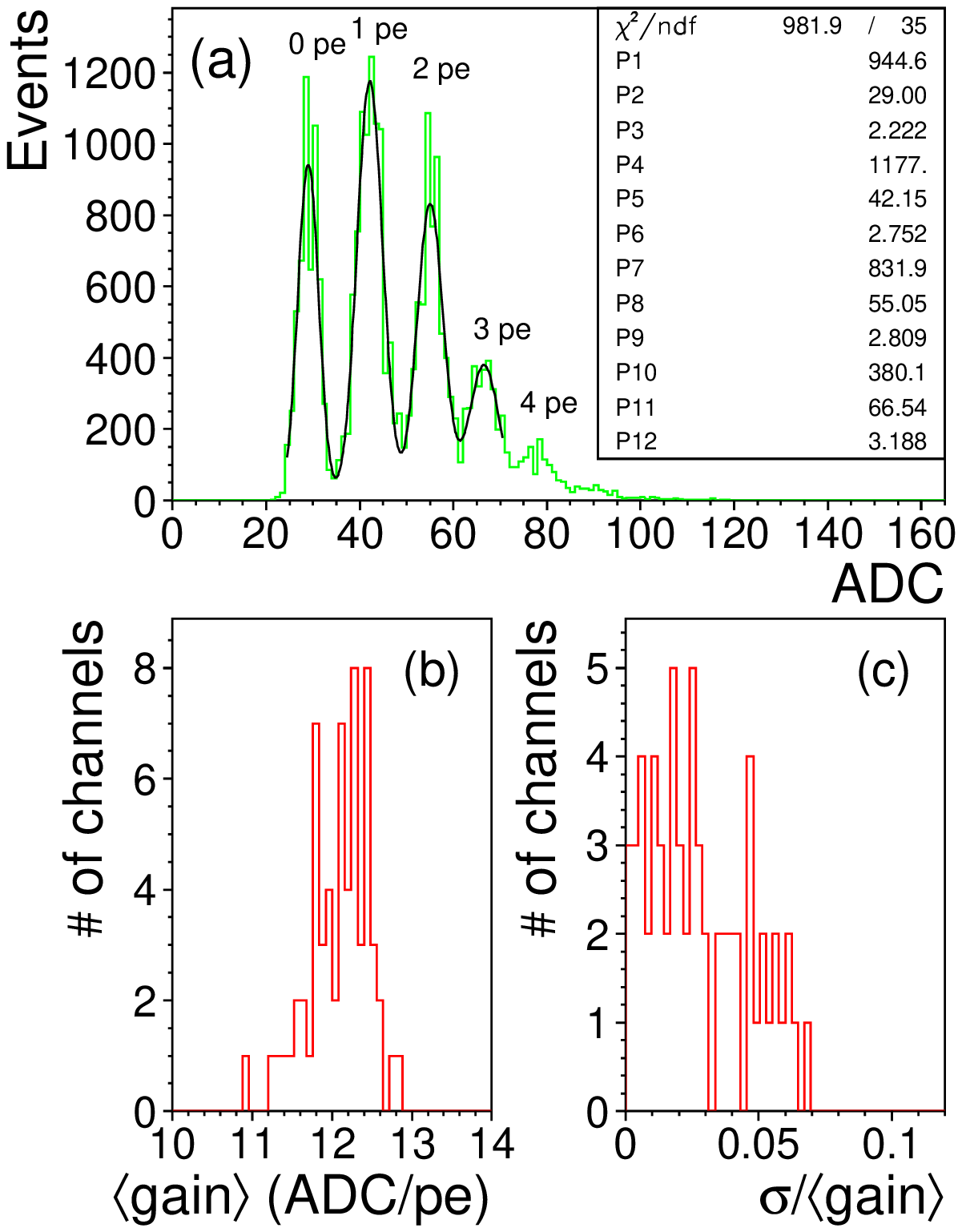}
  \vspace{-.5cm}
  \caption{ A four-Gaussian fit to the LED calibration is shown in (a)
     for a typical VLPC channel.
     The average photoelectron gain over peaks and the 
     corresponding standard deviation are derived.
     The distributions of the average gain and the normalized standard
     deviation for 64 channels of a typical SVX readout are shown 
     in (b) and (c), respectively.
  \label{fig:led4g2} }
\end{figure}                             %----------------------Fig.10

\begin{figure}[b!]                       %----------------------Fig.11
  \vspace{-.5cm}
  \centering
  \includegraphics[width=.65\linewidth] {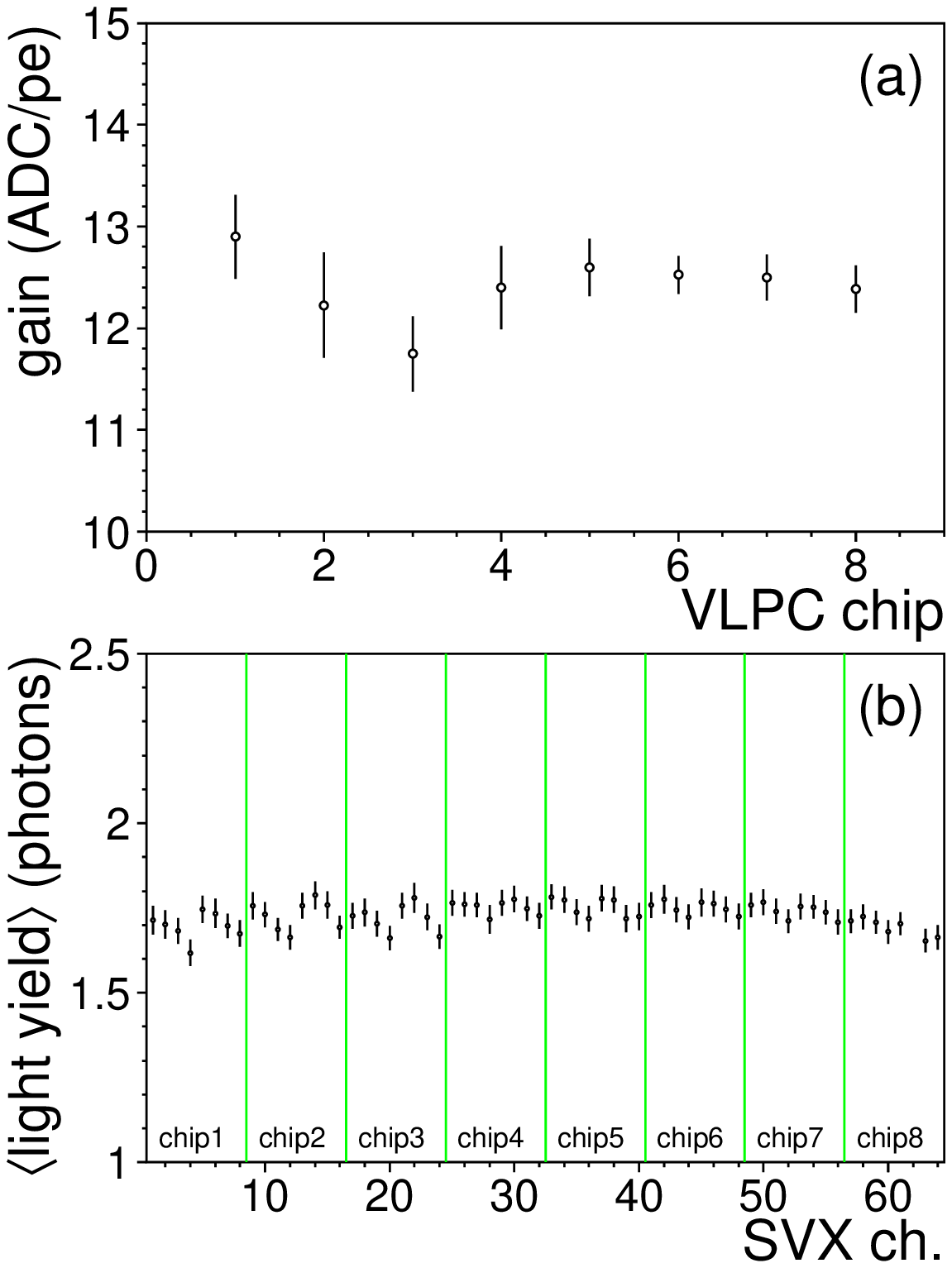}
  \vspace{-.5cm}
  \caption{ (a) The average photoelectron gain for VLPC chips
   (each containing 8 pixels) derived from the LED calibration,
    and  (b) the average light yield taken as the number of
    photons weighted by the Gaussian areas of the fit
    for a typical SVX readout.
  \label{fig:led4g3} }
  \vspace{-.5cm}
\end{figure}                             %----------------------Fig.11

%=============================================================Section 4
\section {VLPC calibration}

The VLPC chips used within a module have similar gain 
and quantum efficiency.
A typical distribution for output from a VLPC pixel (without input)
is shown in Fig.~\ref{fig:led1g}.a, where the pedestal peak 
is fit to a Gaussian function, and the arrow indicates the
$3\sigma$ threshold separating the pedestal from signals
originating from thermal electrons.
The pedestal width is typically 2 to 3 ADC counts, equivalent
typically to $<$0.3 photoelectrons.

The 64 channels on each side of a VLPC module have a common bias 
and the signals are processed by the same SVX chip.
The mean and width of each Gaussian fit to the pedestal 
are plotted in Fig.~\ref{fig:led1g}.b.
The bias voltage is chosen to attain high quantum efficiency and 
low noise;
but the optimum deviates chip-by-chip, as reflected in the
fraction of events that have signals from thermal electrons
(Fig.~\ref{fig:led1g}.c).
The fraction of thermal electrons, typically kept at around 5\%,
provides an indication of proper biasing.

The gain calibration for photoelectrons is performed with 
a LED device.
The LED light is diffused through a bundle of clear fibers 
located above 250 mm long rectangular tubes mounted on the 
warm-end connection to the VLPC.
Each tube is matched to one VLPC module.
Several LED pulser configurations were used in the calibration 
procedure to obtain distributions with varying numbers 
of photoelectrons.
A typical distribution for a moderate light yield,
fitted to a distribution of four Gaussians,
is shown in Fig.~\ref{fig:led4g2}.a.
The clearly separated peaks correspond to the pedestal and 
observation of up to five photoelectrons.

The peak-to-peak uniformity of photoelectron gain for each 
channel is obtained from the four Gaussians of the fit.
The photoelectron gains are derived from the differences 
between adjacent Gaussian means.
The average gain and the standard deviation are calculated.
Results from the 64 channels of a typical SVX readout
are presented in Figs.~\ref{fig:led4g2}.b and c, respectively.
The VLPC pixels have compatible gains of around 12 ADC counts per
photoelectron, and the standard deviations are all within 5\%.

The photoelectron gains within pixels of the same VLPC chip are uniform
to within 2\% and are grouped together during the calibration phase.
The means with standard deviations are plotted in 
Fig.~\ref{fig:led4g3}.a for the 8 chips of a typical SVX readout.
The light yield from the calibration is calculated 
from the number of photoelectrons, weighted by the Gaussian areas
of the fit, and plotted for each pixel in Fig.~\ref{fig:led4g3}.b.
The yield is reasonably uniform over all pixels.
The incident LED light has slightly higher intensity 
in the central region of the VLPC module, and the deviation
seen every four channels corresponds to the positions of pixels
located between the center and the edge of the module.

\begin{figure}[b!]                     %----------------------Fig.12
  \vspace{-.5cm}
  \centering
  \includegraphics[width=.60\linewidth] {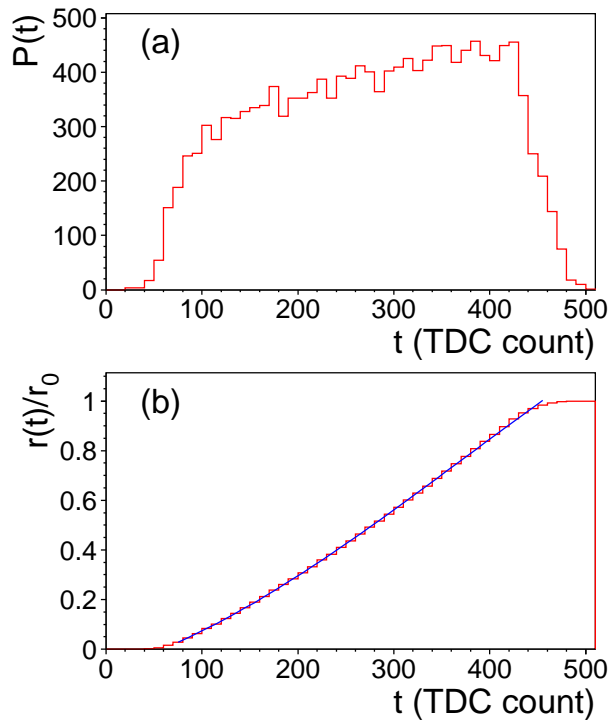}
  \vspace{-.5cm}
  \caption{(a) A typical TDC spectrum for a PDT channel, and 
           (b) the integrated event fraction with a fit
               to a third order polynomial.
  \label{fig:mpdt_cali} }
  \vspace{-.5cm}
\end{figure}                           %----------------------Fig.12

%=============================================================Section 5
\section {PDT track reconstruction and cosmic ray selection}

The PDT data that provide reference tracking are processed
independently of the VLPC data stream.
A typical PDT drift time distribution 
is shown in Fig.~\ref{fig:mpdt_cali}.a.
The cosmic ray muons are distributed approximately uniformly
across the tube.
The higher event fraction at larger drift times indicates a slower 
drift speed for ionization charge produced at large distances
from the anode.
The integrated event fraction, $\int P(t)$, as a function of drift time,
shown in Fig.~\ref{fig:mpdt_cali}.b, is fit to a third order polynomial,
and is used for calculating the time-to-distance function.
For a given drift time $t$, the distance to the anode is 
\begin{equation}
  r(t) = r_0  \displaystyle \int _0^{t} P(t') \; dt',
\end{equation}
where $r_0$ is the tube radius.
The position where the track traversed the PDT
is calculated, first, assuming a vertical track,
and iterated after the track fitting is performed 
to correct for the incident angle.

\begin{figure}[t!]                    %----------------------Fig.13
  \vspace{-1.cm}
  \centering
  \includegraphics[width=.60\linewidth] {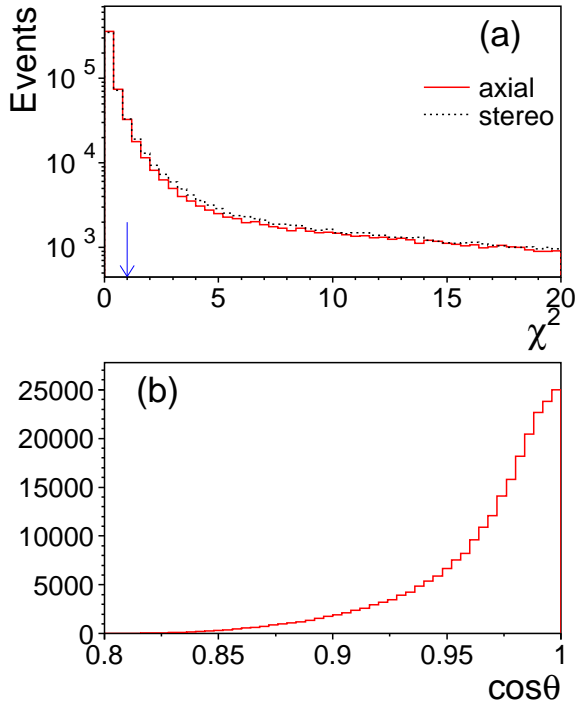}
  \vspace{-.8cm}
  \caption{(a) $\chi^2$ distributions for unweighted linear fits
           to axial and stereo PDT hits,
           and (b) the $\cos \theta$ distribution for 3-D tracks.
  \label{fig:mpdt} }
  \vspace{.5cm}
\end{figure}                          %----------------------Fig.13

The PDT diameter is 30 mm, the offset between adjacent tubes 
is 25 mm in the horizontal $y$-axis,
and 20 mm in the vertical $z$-axis.
A large fraction of the tracks passes through a 
PDT layer hitting just a single tube, which introduces
a two-fold ($\pm y$) ambiguity for the drift direction.
Each direction is tried, and the one with the 
best $\chi^2$ value for the linear track is selected.

An unweighted linear fit \cite{LFIT} is performed to the PDT hit
positions in $y$ versus the anode position in $z$.
The $\chi^2$ distributions obtained for fits 
to the axial and stereo layers are plotted in 
Fig.~\ref{fig:mpdt}.a.
Events containing a single track are selected
if both $\chi^2$ values are smaller than one.
The selected tracks on average have four hits, and
the $\cos \theta$ distribution for the polar angle $\theta$ 
to the $z$-axis is plotted in Fig.~\ref{fig:mpdt}.b.
The $y$-coordinate of the three-dimensional (3-D)
reference track is derived from the fit to axial PDT hits,
and the $x$-coordinate comes from the fit to stereo hits.

\begin{figure}[b!]                    %----------------------Fig.14
  \vspace{-1.cm}
  \centering
  \includegraphics[width=.60\linewidth] {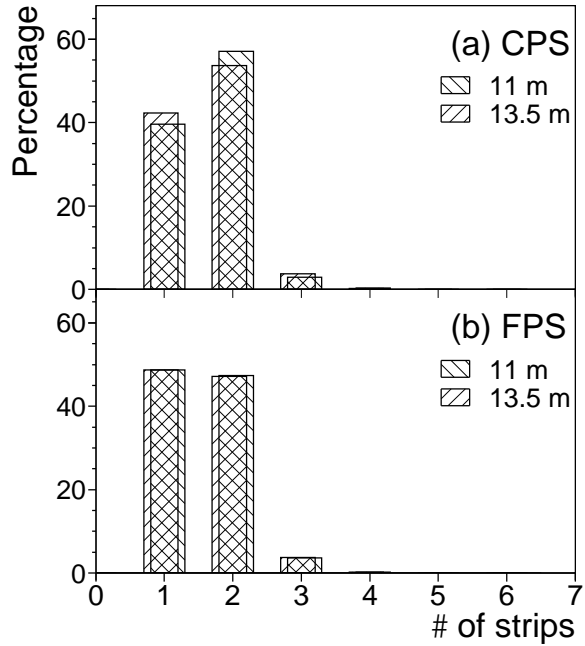}
  \vspace{-.5cm}
  \caption{ Distributions in the number of strips in a cluster,
            for (a) the CPS and (b) the FPS modules for
            11 m and 13.5 m clear fibers.
  \label{fig:qclre_nclr} }
\end{figure}                         %----------------------Fig.14

\begin{figure}[t!]                    %----------------------Fig.15
  \vspace{-.5cm}
  \centering
  \includegraphics[width=.58\linewidth] {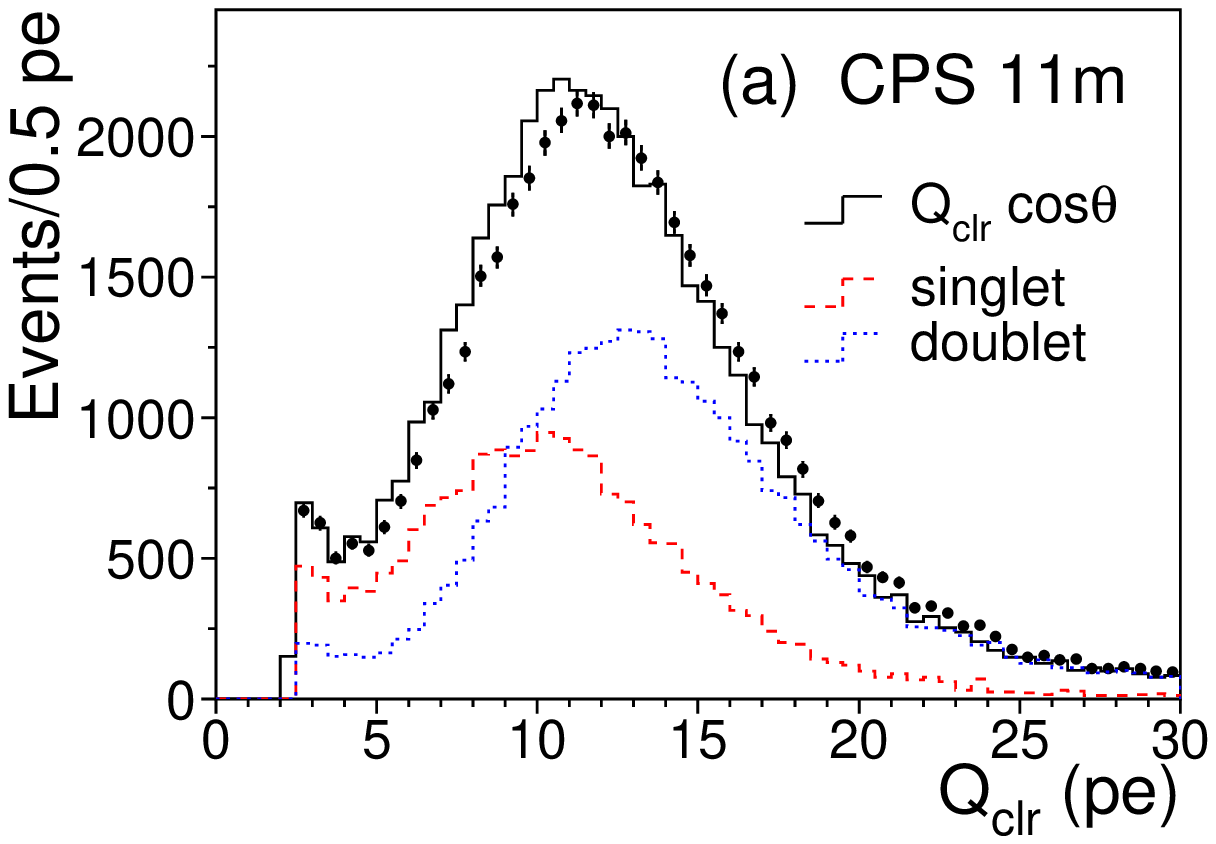}

  \vspace{-1.2cm}
  \includegraphics[width=.58\linewidth] {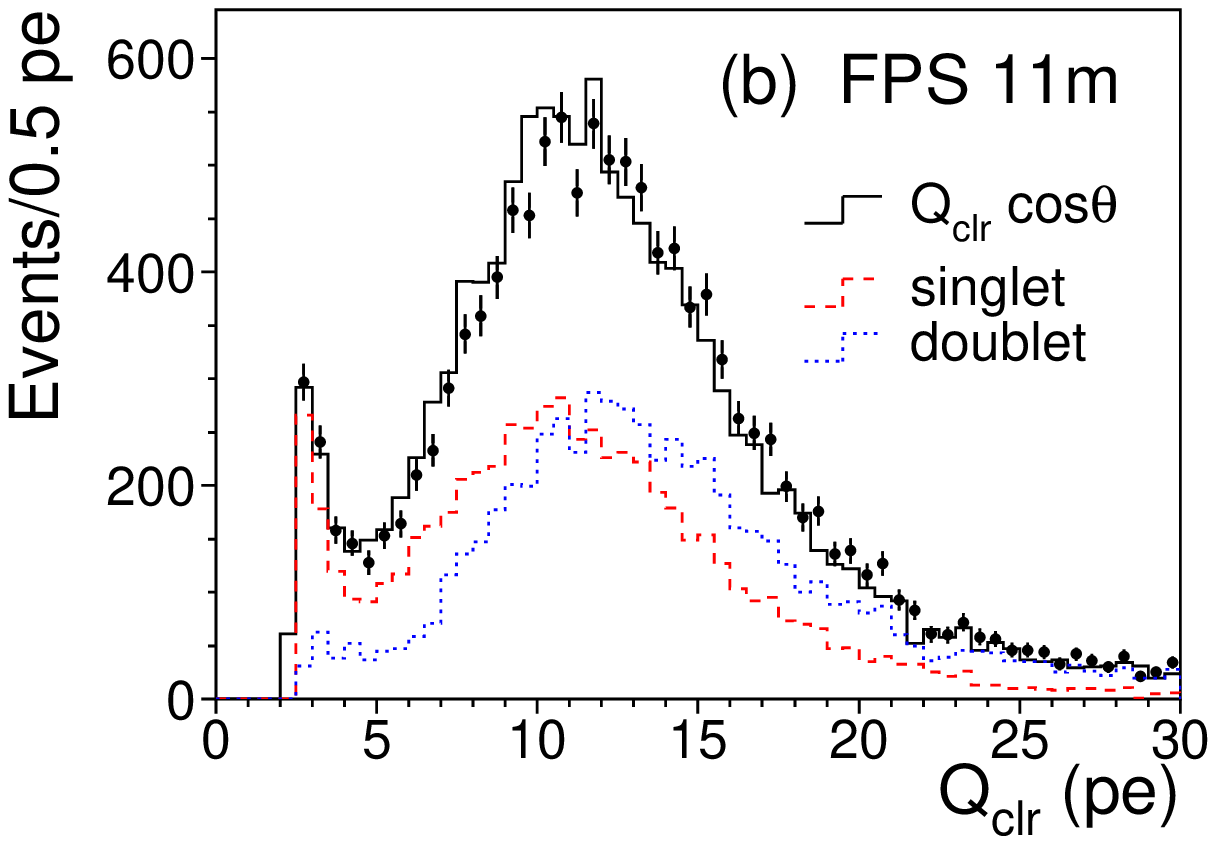}
  \vspace{-.5cm}
  \caption{The light yield for minimum ionizing particles for
           (a) CPS-south and 
           (b) FPS-$u$, collected with 11 m clear fibers.
           Distributions for singlet and doublet clusters are shown as
           dashed and dotted lines, respectively. 
           The light yield corrected for $\cos \theta$ is shown in 
           the solid line.
  \label{fig:qclre_11} }
  \vspace{.5cm}
\end{figure}                         %----------------------Fig.15

\begin{figure}[p]                    %----------------------Fig.16
  \vspace{-2.0cm}
  \centering
  \includegraphics[width=.58\linewidth] {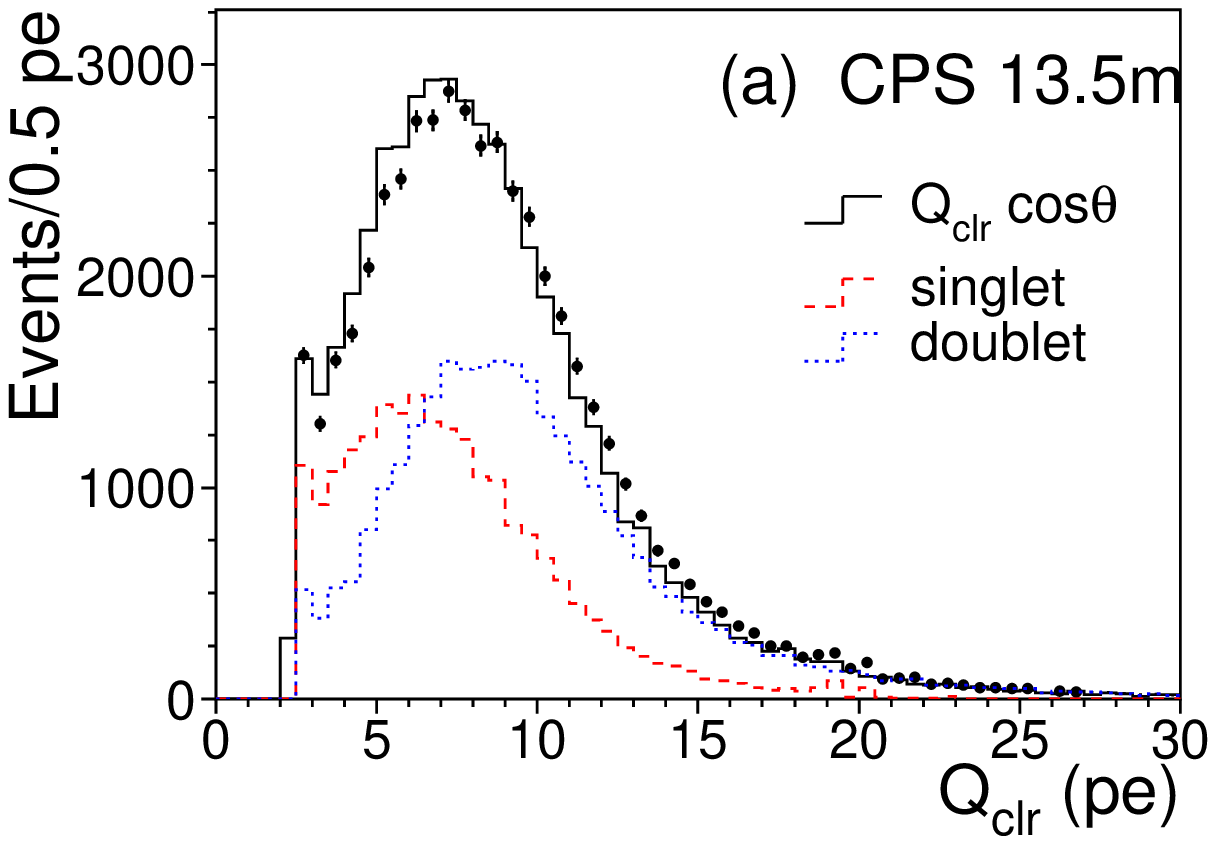}

  \vspace{-1.2cm}
  \includegraphics[width=.58\linewidth] {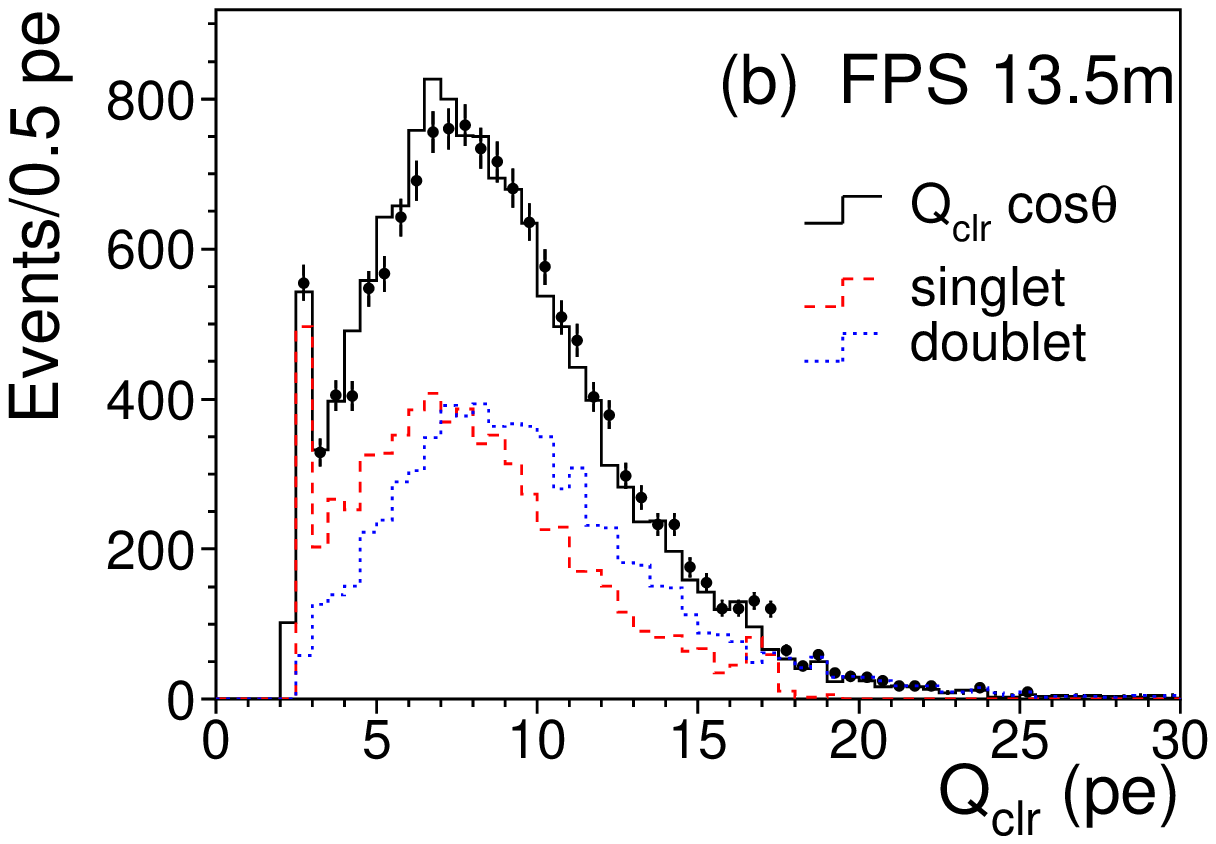}
  \vspace{-.5cm}
  \caption{The light yield for minimum ionizing particles for
           (a) CPS-north and 
           (b) FPS-$v$, collected with 13.5 m clear fibers.
           Distributions for singlet and doublet clusters are shown as
           dashed and dotted lines, respectively. 
           The light yield corrected for $\cos \theta$ is shown in 
           the solid line.
  \label{fig:qclre_13} }
% \end{figure}                         %----------------------Fig.16
% \begin{figure}[b!]                   %----------------------Fig.17
  \vspace{-1.cm}
  \centering
  \includegraphics[width=.58\linewidth] {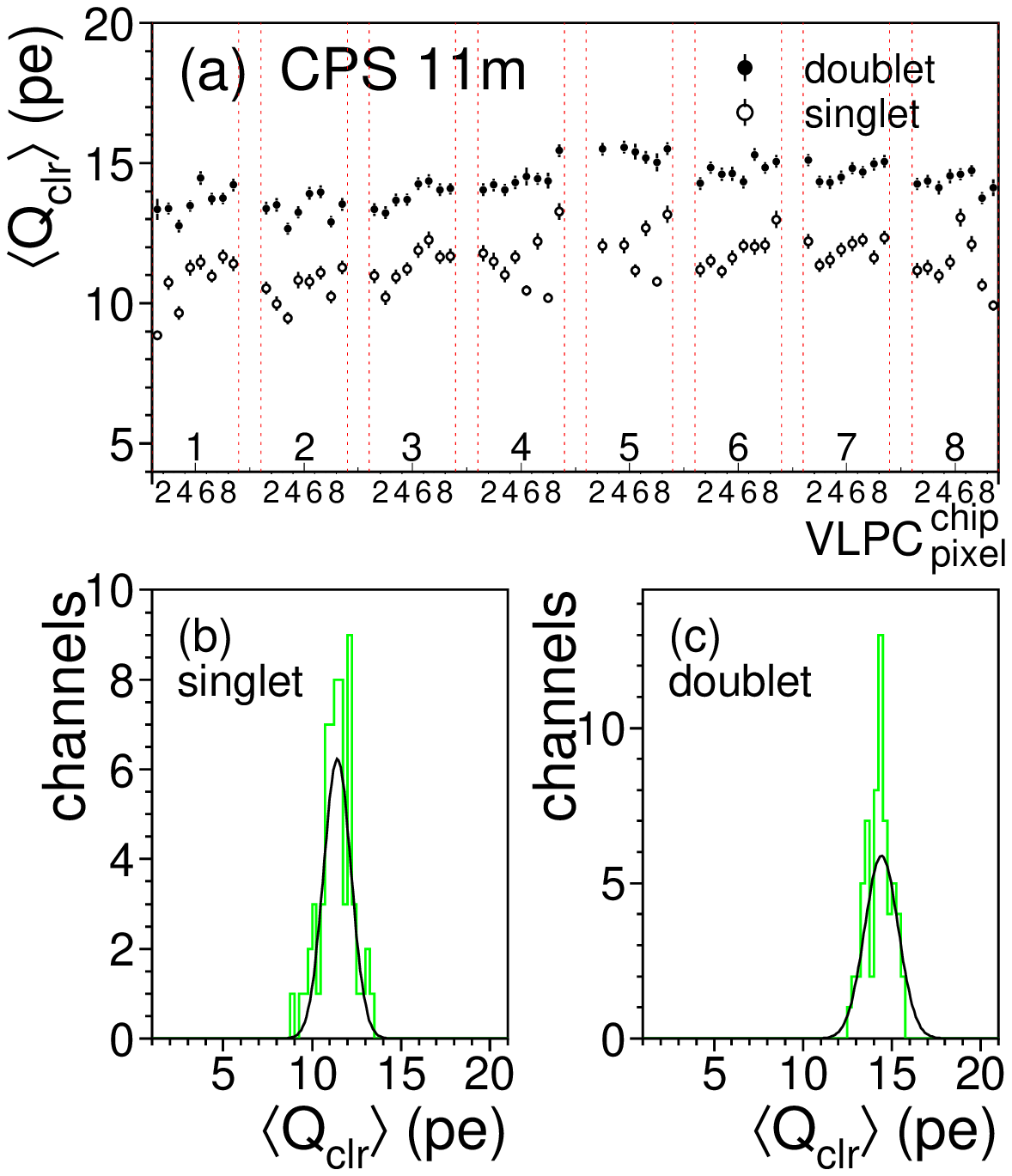}
  \vspace{-.5cm}
  \caption{The average light yield for MIPs 
           observed by the CPS-south segment with 11 m clear fibers.
           In (a) is the average light yield versus strip number,
           where the latter is labeled by the corresponding 
           VLPC pixel identity.
           The distributions of means are plotted in 
           (b) for singlet and (c) for doublet clusters, with 
           Gaussian fits shown by the curves.
  \label{fig:qclre_mcps_12} }
\end{figure}                         %----------------------Fig.17

%=============================================================Section 6
\section {Preshower MIP light yield}

%===============================================
\subsection {Strip cluster }

The triangular scintillator strips are stacked 
such that the fraction of vertical tracks traversing two strips
is 67\% for the CPS and 62\% for the FPS.
The SVX outputs are in units of ADC counts, and are first converted 
to number of photoelectrons (pe).  
The SVX channel numbers are mapped to the sequence of the strip layout.
A strip cluster for a traversing track is reconstructed requiring:
\begin{list}{}{\setlength{\itemsep}{0pt}
               \setlength{\parsep}{0pt}
               \setlength{\topsep}{0pt} }
  \item[i.] a channel with maximum light yield larger
            than 2 pe,
  \item[ii.] adjacent channels with descending values of
             signal above a cutoff threshold of 0.7 pe,
  \item[iii.] and a total cluster light yield, $Q_{clr}$,
              to be larger than 3 pe.
\end{list}

The distributions in the number of strips contained in a cluster 
are shown in Fig.~\ref{fig:qclre_nclr}.
The cluster light yield observed for cosmic ray muons 
is uniform over the readout channels.
The light-yield spectra obtained with 11 m clear fibers are 
plotted in Figs.~\ref{fig:qclre_11}.a and b,
for the CPS-south segment and the FPS-$u$ layer, respectively.
Similar spectra obtained with the 13.5 m clear fibers are plotted in 
Fig.~\ref{fig:qclre_13}.
Contributions from single (singlet) and two-strip (doublet) 
clusters are shown by the dashed and dotted lines, respectively.
The fraction of doublets with 11 m clear fibers
is 61\% for the CPS and 53\% for the FPS, 
which is about 10\% lower than the geometric fractions 
caused by the low light yield and selection cutoffs
for muons traversing the corners of the strips.
The average polar angle of tracks is about $15^\circ$
(Fig.~\ref{fig:mpdt}.b), corresponding roughly to a 4\% longer
distance in the scintillator than along the vertical direction.
The light yields, normalized to vertical tracks, are shown by
the solid lines.

The average light-yield profile versus the fiber number of
the cluster center, for the CPS with 11 m clear fibers,
is shown in Fig.~\ref{fig:qclre_mcps_12}.a.
The light-yield uniformity depends on the quality
of scintillator strips, WLS and clear fibers,
and the quantum efficiency of VLPC chips at the same bias voltage.
The light yield tends to be correlated with VLPC chip in levels,
which is a indication for the difference in quantum efficiency.
Singlet clusters have lower light yield, in part due to
the smaller depth of scintillator.
The distributions of average light yield for singlets and doublets are
shown in Figs.~\ref{fig:qclre_mcps_12}.b and c, respectively, 
and fitted to Gaussians.
Results of the fits for the CPS and the FPS with 11 m and 13.5 m
clear fibers are presented in Table \ref{tab:MIP}.
The standard deviations obtained are in general less than 10\% 
of the mean.
The singlet light yields are about 20\% lower than those 
for doublets.

The light collection is improved with optical grease applied 
at the end-faces of the clear fibers.
The mean light yield was compared to that in a setup without grease.
Applying Bicron BC-630 optical grease at one end of the 
clear fibers increases the mean yield by about 7\%.
When grease is applied at both ends, the mean increases by 14\%.
Light attenuation along the scintillator strips is studied for
the CPS module as a function of PDT track position in $x$.
The loss due to attenuation is about 10\% over the 1206 mm strip length.

\begin{table}[t!]                   %-------------------------Table 1
  \begin{center}
   \begin{tabular}{l|c|cc|cc} 
  \hline
  \hline  Detector  &  clear-fiber  & \multicolumn{2}{c|}{singlet} &
                                      \multicolumn{2}{c}{doublet} \\
                    &   length (m)  &  mean (pe) & $\sigma$/mean
                                    &  mean (pe) & $\sigma$/mean  \\
  \hline  CPS-south &  11   & $11.4\pm0.1 $  & $7.9\pm1.5$\% 
                            & $14.3\pm0.2 $  & $5.0\pm1.3$\% \\
          FPS-$u$   &  11   & $11.8\pm0.3 $  & $11.7\pm3.3$\% 
                            & $14.0\pm0.3 $  & $9.5\pm2.3$\% \\
          CPS-north &  13.5 & $ 8.0\pm0.1 $  & $6.7\pm0.8$\%
                            & $10.0\pm0.2 $  & $7.2\pm2.1$\% \\
          FPS-$v$   &  13.5 & $ 8.5\pm0.2 $  & $9.2\pm3.0$\%
                            & $10.4\pm0.2 $  & $7.7\pm2.1$\% \\
  \hline
  \hline
    \end{tabular}
    \caption{ The mean values and standard deviations for light yield,
              with statistical errors from Gaussian fits 
              to distributions for singlets and doublets
           (cf.\ Figs.~\ref{fig:qclre_mcps_12}.b and c, respectively).
    \label{tab:MIP}  }
  \end{center}
  \vspace{.5cm}
\end{table}                       %--------------------------Table 1

\begin{figure}[b!]                    %----------------------Fig.18
  \vspace{-1.cm}
  \centering
  \includegraphics[width=.8\linewidth] {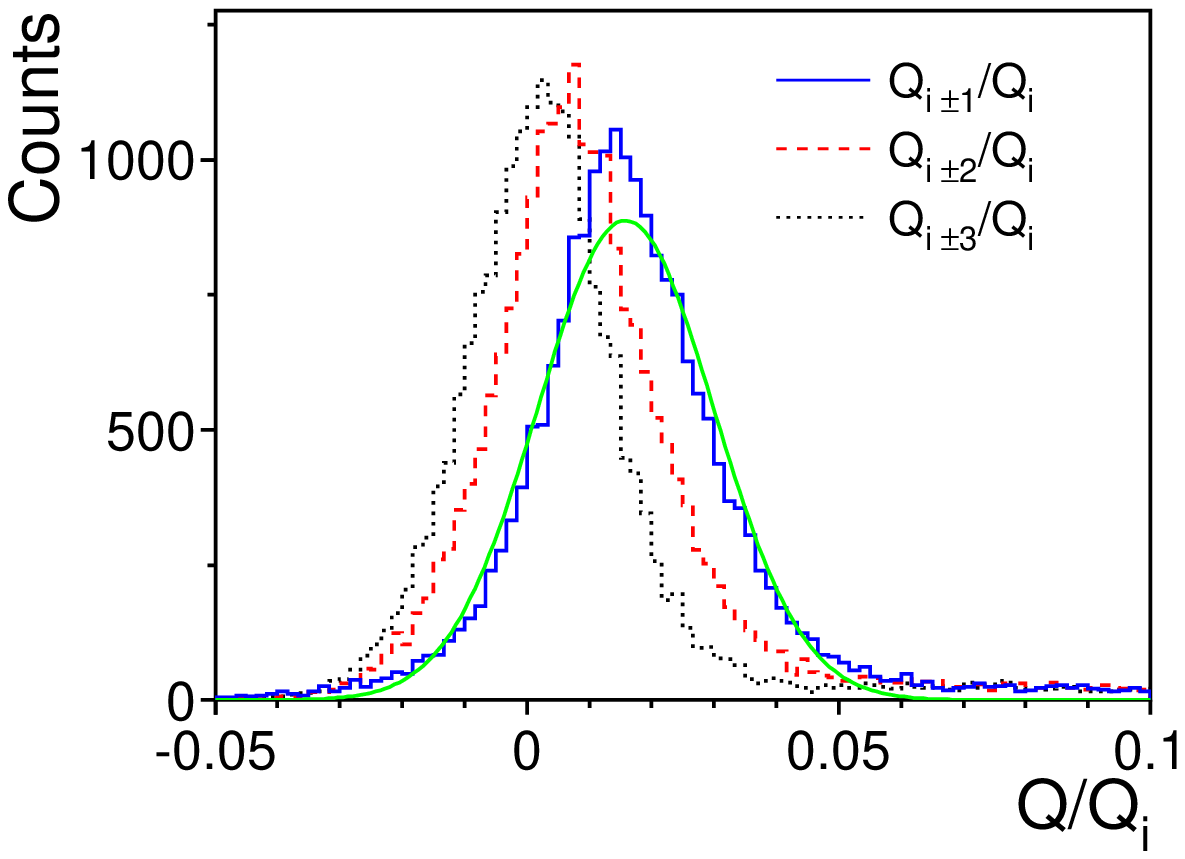}
  \vspace{-.6cm}
  \caption{Cross talk in the electronics and 
           between traces on the flex cable.
           For a selected signal channel-$i$, plotted is
           the normalized charge found one ($i\pm1$),
           two ($i\pm2$) and three ($i\pm3$) channels 
           away from the signal channel.
  \label{fig:xtalk_elec} }
  \vspace{-.5cm}
\end{figure}                          %----------------------Fig.18

\begin{figure}[t!]                    %----------------------Fig.19
  \vspace{-1.cm}
  \centering
  \includegraphics[width=.8\linewidth] {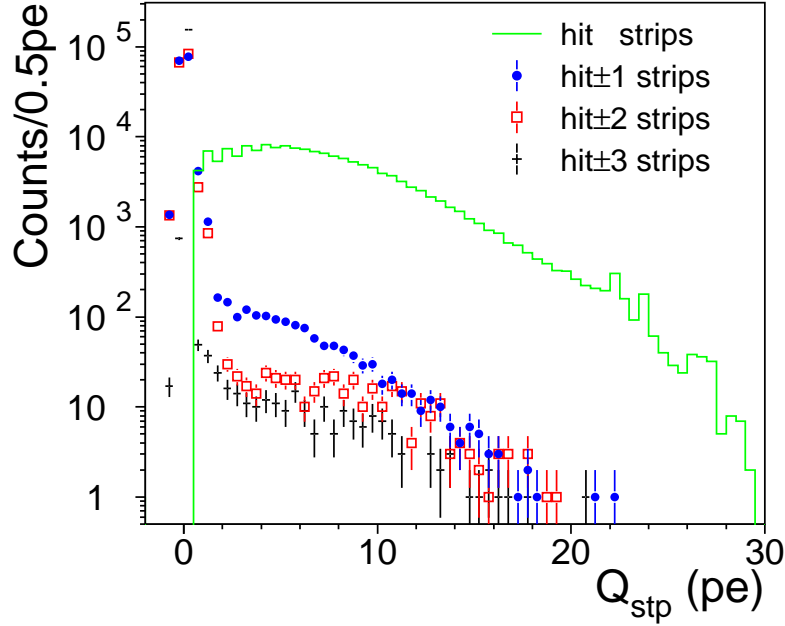}
  \vspace{-.6cm}
  \caption{Light-yield spectra of scintillator strips 
           one (hit$\pm$1), two (hit$\pm$2) and three (hit$\pm$3)
           strips away from the two central hit-strips.
  \label{fig:xtalk_scin} }
  \vspace{.5cm}
\end{figure}                          %----------------------Fig.19

%====================================================
\subsection {Cross talk and delta-rays }

Cross talk in the electronics and between the traces
on the flex cable was investigated for channels with light yield 
larger than 10 photoelectrons.
The charge seen in adjacent channels, normalized to the one selected
($Q_{i\pm1}/Q_i$), is plotted in Fig.~\ref{fig:xtalk_elec}.
A Gaussian fit to the distribution gives a mean of 1.6\% with 
a standard deviation of 1.4\%.
The clear fiber connections are arranged such that
adjacent scintillator strips are read out through non-adjacent
electronic channels,
thereby decoupling cross talk between adjacent electronic
channels from light leakage or delta-rays crossing adjacent strips.

The same analysis was made for channels further away from
the selected one.
The distributions for cross talk to two ($Q_{i\pm2}/Q_i$) 
and three channels ($Q_{i\pm3}/Q_i$) away from the central one, 
have Gaussian means of 0.7\% and 0.2\%, respectively,
and standard deviations of 1.4\% and 1.2\%.
The magnitude of the cross talk is observed to be
similar for all channels.
The flex cable is designed for minimum heat transfer
from the cold end in the VLPC cassette to the front-end board
at room temperature.
Cross talk between the traces is anticipated, and the small value
observed is within design tolerance.

Light leakage between scintillator strips and clear fibers
as well as delta-ray contributions were investigated using
clusters containing two or more strips.
The two adjacent strips with the most light deposition are
denoted as the {\it hit-strips} which the muon traversed.
The light-yield spectra for individual strips ($Q_{stp}$)
are plotted in Fig.~\ref{fig:xtalk_scin}.
At a threshold of 2 photoelectrons, the cumulative event fraction
for the strips adjacent to the hit-strips (hit$\pm1$) is 0.85\%,
indicating a negligible light leakage between strips and clear fibers.
Occasionally the adjacent strips can have large
light deposition, and the event fraction decreases
slowly with the distance to the hit-strips.
The cumulative event fractions (above 2 photoelectrons) observed in
strips two and three strips away from the hit-strips are 0.25\% 
and 0.15\%.
These are most likely events containing delta-rays generated upstream.
These event fractions for different CPS and FPS modules are 
consistent with each other.

\begin{figure}[b!]                     %------------------Fig.20
  \vspace{-.5cm}
  \centering
  \includegraphics[width=.60\linewidth] {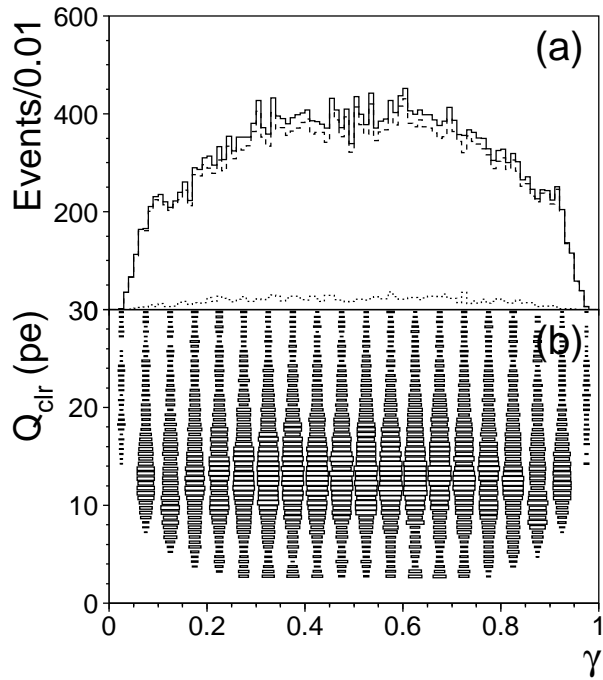} 
  \vspace{-.5cm}
  \caption{
      Distributions of (a) $\gamma$ and (b) light yield versus $\gamma$,
      for multiple strip clusters of the CPS with 11 m clear fibers.
      The dashed and dotted lines in (a) correspond to clusters with 
      two or more than two strips, respectively.
  \label{fig:eta_1} }
  \vspace{-.5cm}
\end{figure}                           %------------------Fig.20

\begin{figure}[b!]                      %------------------Fig.21
% \vspace{-1.cm}
  \centering
  \includegraphics[width=.65\linewidth] {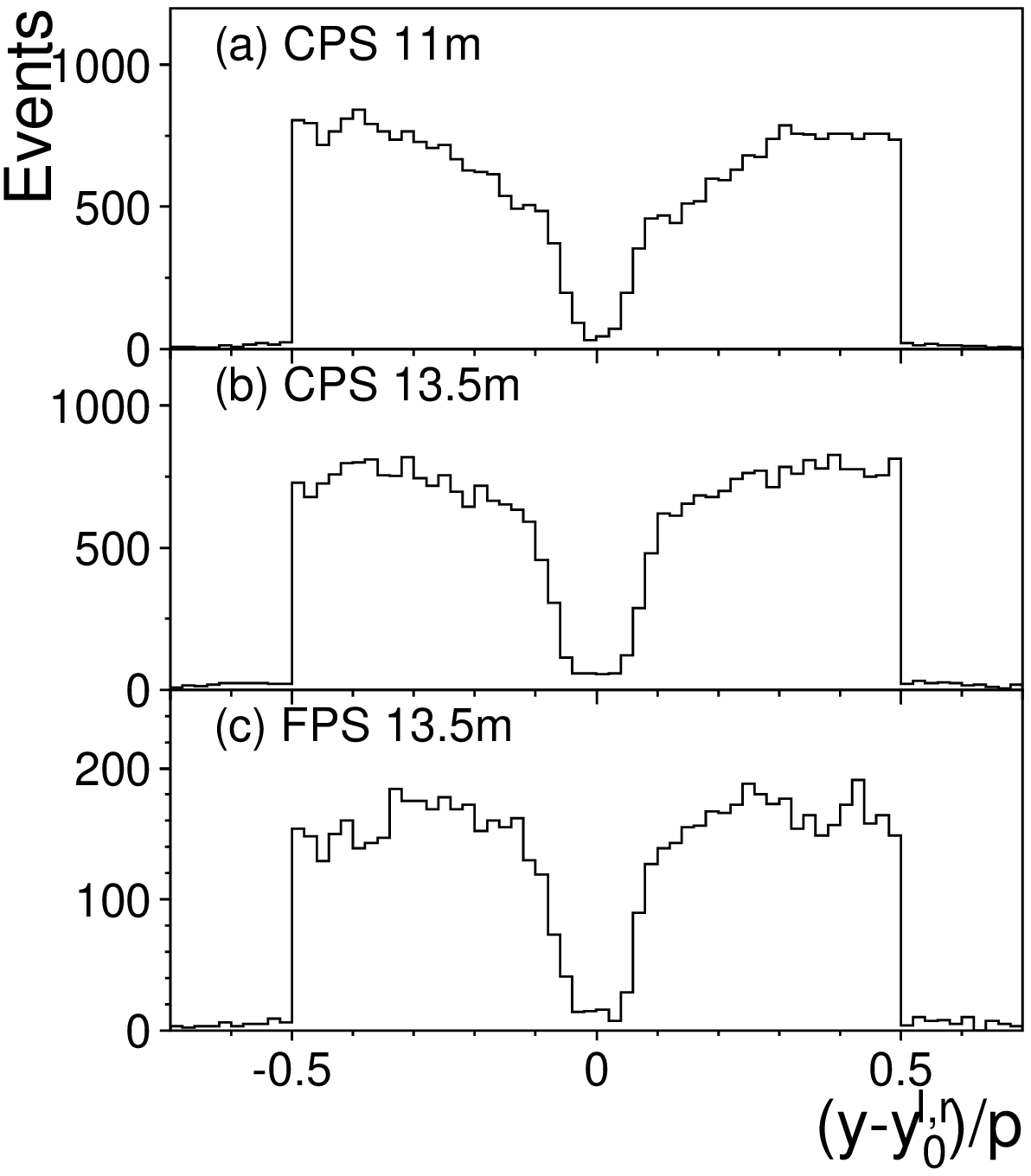}
  \vspace{-.2cm}
  \caption{Cluster position to the closest 
           strip center ($y_0^{l,r}$) in units of strip pitch
           for CPS with
           (a) 11 m, and (b) 13.5 m clear fibers, and c) 
           FPS with 13.5 m clear fibers.
  \label{fig:eta_3} }
  \vspace{-.5cm}
\end{figure}                            %------------------Fig.21

\begin{figure}[t!]                     %------------------Fig.22
  \vspace{-.8cm}
  \centering
  \includegraphics[width=.65\linewidth] {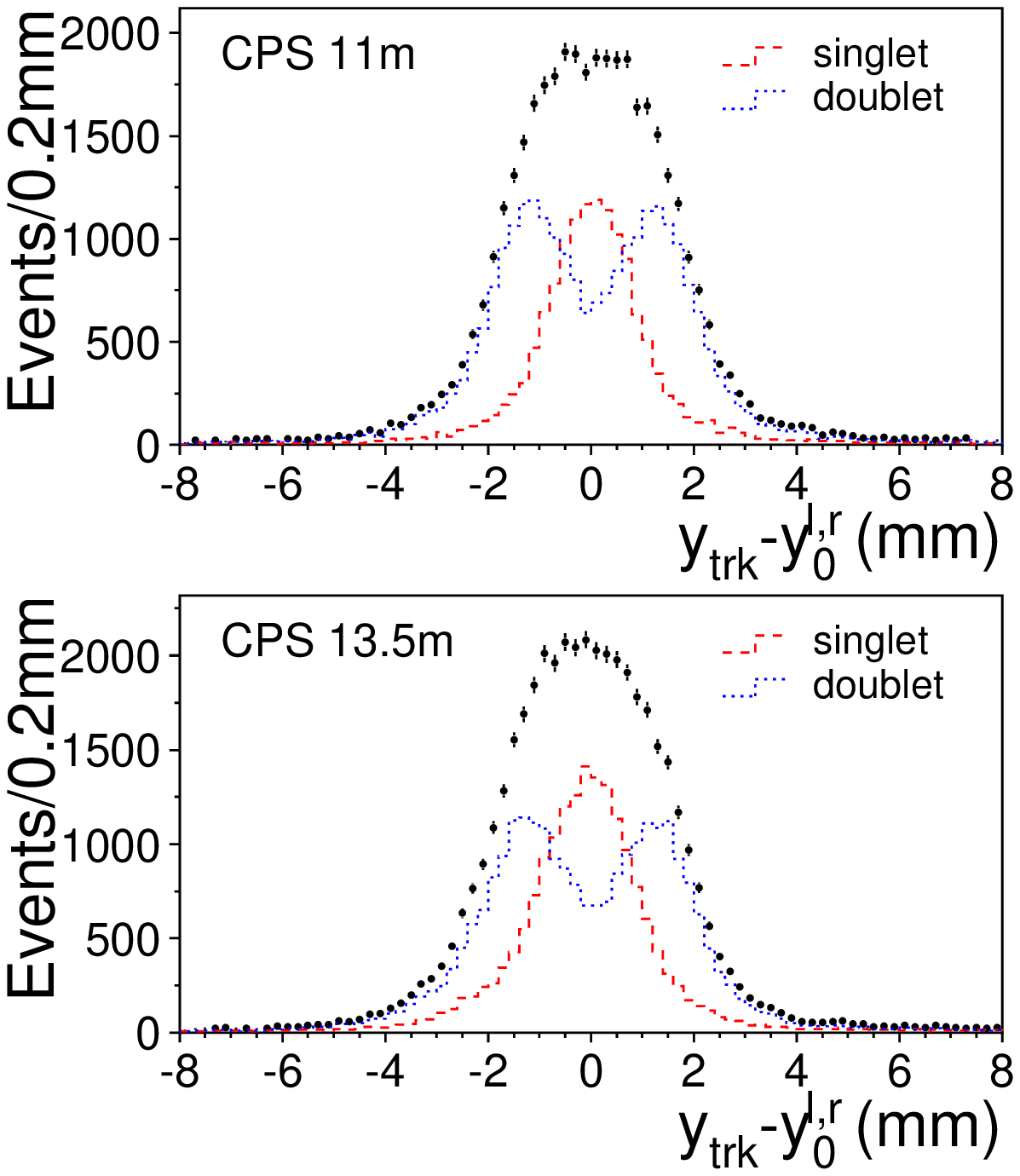}
  \vspace{-.2cm}
  \caption{PDT track position ($y_{trk}$) relative to the center 
           of the strip ($y_0^{l,r}$) where a cluster is found, 
           for CPS with (a) 11 m and (b) 13.5 m 
           clear fibers, respectively.
           Constituent distributions for single (dashed)
           and doublet clusters (dotted) are also shown.
  \label{fig:eta_4} }
  \vspace{.5cm}
\end{figure}                            %------------------Fig.22

%=============================================================Section 7
\section {Spatial resolution }

%====================================================
\subsection {Strip cluster position }

The triangular shape of the scintillator strips is a 
convenient configuration for reconstructing the position
of a particle that passes through two strips.
The distance traversed by the track in each strip has a linear 
correspondence to the incident position.
The cluster position can therefore be calculated by using
the charge-weighted mean of the strip centers.
The light sharing between two neighboring strips is expressed 
by the ratio
\begin{equation}
  \gamma =  \frac{Q_l} {Q_l + Q_r},
\end{equation}
where $Q_r$ and $Q_l$ are the strip signals of the right and
left channels of a doublet cluster.
If a cluster contains more than two strips, possibly from a delta-ray
or thermal noise,
then $Q_r$ and $Q_l$  are the sum of channels on the right and left 
of the charge weighted mean.

The distribution in $\gamma$ and total yield in cluster light
as a function of $\gamma$ are plotted in 
Fig.~\ref{fig:eta_1} for the CPS with 11 m clear fibers.
Muon tracks are assumed to be distributed uniformly across a strip.
In the ideal case, one would anticipate a uniform distribution
in $\gamma$. 
The round-offs near $\gamma$=0 and 1 correspond to regions of low 
light yield and to the individual strip threshold of 0.7 photoelectrons 
that removes tracks traversing a corner of a strip.
Clusters of more than two strips account for about 6\% 
of the total events.
The contamination from thermal electrons can be suppressed by 
a higher cutoff threshold, 
which produces a wider gap at $\gamma$=0 and 1,
and the tradeoff of having more singlet clusters.

The cluster position for doublets is derived for the light sharing by 
\begin{equation}
  y(\gamma) = y_0 + p \cdot \gamma, 
  \label{eq:y}
\end{equation}
where $p$ is the strip pitch and $y_0$ the center of the right-hand
strip.
The distributions of cluster position relative to
the closest strip center ($y_0^{l,r}$), in units of the strip pitch,
are shown in Fig.~\ref{fig:eta_3}.
Singlet clusters, which would fall at $y \sim y_0^{l,r}$ in
this algorithm, are not plotted.
The CPS and FPS modules with 13.5 m clear fibers have lower 
light yields, and the strip cutoff threshold 
therefore creates a wider gap at $y$ near $y_0^{l,r}$.
The shape reflects the combined effects of the geometry
and the Poisson statistics of low light yield near a strip corner.

Spatial resolution is investigated with respect to 
hit positions given by the PDT track.
The strip cluster position is corrected for alignment relative
to the PDTs, which includes the offsets and rotations 
of strip positions.
The preshower strips are mounted on cylindrical platforms, and 
the strip pitch and $z$ position are therefore corrected accordingly.
The dominant systematic uncertainties arise from
the effects of multiple scattering and strip alignment.
The effects are seen in Fig.~\ref{fig:eta_4}
which shows the difference between the PDT track position 
and the center of the strip where a cluster is found.
A uniform distribution would be expected for ideal resolution
with the doublet component of the distribution looking similar 
to Fig.~\ref{fig:eta_3} and the singlet component filling
the central gap.

\begin{table}[b!]                   %-------------------------Table 2
  \vspace{.5cm}
  \begin{center}                    % was 00 p031, new 00p78
   \begin{tabular}{ll|cccc} 
  \hline
  \hline      &    & $\chi^2$/ndf &  $\sigma_n$ (mm) & $\sigma_w$ (mm)
                                  &  $\int g_n / \int (g_n +g_w)$  \\
  \hline
    CPS (11m) & singlet &  53/32  &  $0.72\pm0.02$  & $1.78 \pm 0.05$  
                                                    & $0.69 \pm 0.04$ \\
    CPS (11m) & doublet &  28/30  &  $0.67\pm0.01$  & $1.67 \pm 0.05$  
                                                    & $0.63 \pm 0.03$ \\
    CPS (11m) & combined&  32/30  &  $0.69\pm0.01$  & $1.69 \pm 0.05$  
                                                    & $0.64 \pm 0.02$ \\
  \hline
  CPS (13.5m) & singlet &  36/34  &  $0.75\pm0.02$  & $1.63 \pm 0.04$  
                                                    & $0.56 \pm 0.03$ \\
  CPS (13.5m) & doublet &  48/30  &  $0.72\pm0.02$  & $1.64 \pm 0.05$  
                                                    & $0.59 \pm 0.03$ \\
  CPS (13.5m) & combined&  54/32  &  $0.73\pm0.01$  & $1.64 \pm 0.03$  
                                                    & $0.58 \pm 0.02$ \\
  \hline
  FPS (13.5m) & singlet &  37/42  &  $0.83\pm0.02$  & $2.91 \pm 0.20$  
                                                    & $0.71 \pm 0.03$ \\
  FPS (13.5m) & doublet &  43/38  &  $0.75\pm0.02$  & $2.32 \pm 0.12$  
                                                    & $0.66 \pm 0.04$ \\
  FPS (13.5m) & combined&  37/40  &  $0.80\pm0.01$  & $2.66 \pm 0.12$  
                                                    & $0.69 \pm 0.03$ \\
  \hline
  \hline
    \end{tabular}
    \caption{ Parameters of double-Gaussian fits to the residuals 
              of CPS and FPS with 11 m and 13.5 m clear fibers.
              Listed are widths for the narrow core ($g_n$),
              the wide ($g_w$) Gaussians, and the fraction 
              covered by the narrow Gaussian
              ($\int g_n / \int (g_n +g_w)$).
              The errors are statistical.
    \label{tab:resi}  }
  \end{center}
  \vspace{-.5cm}
\end{table}                       %--------------------------Table 2

\begin{figure}[p]                     %----------------------Fig.23
  \vspace{-2.cm}
  \centering
  \includegraphics[width=.75\linewidth] {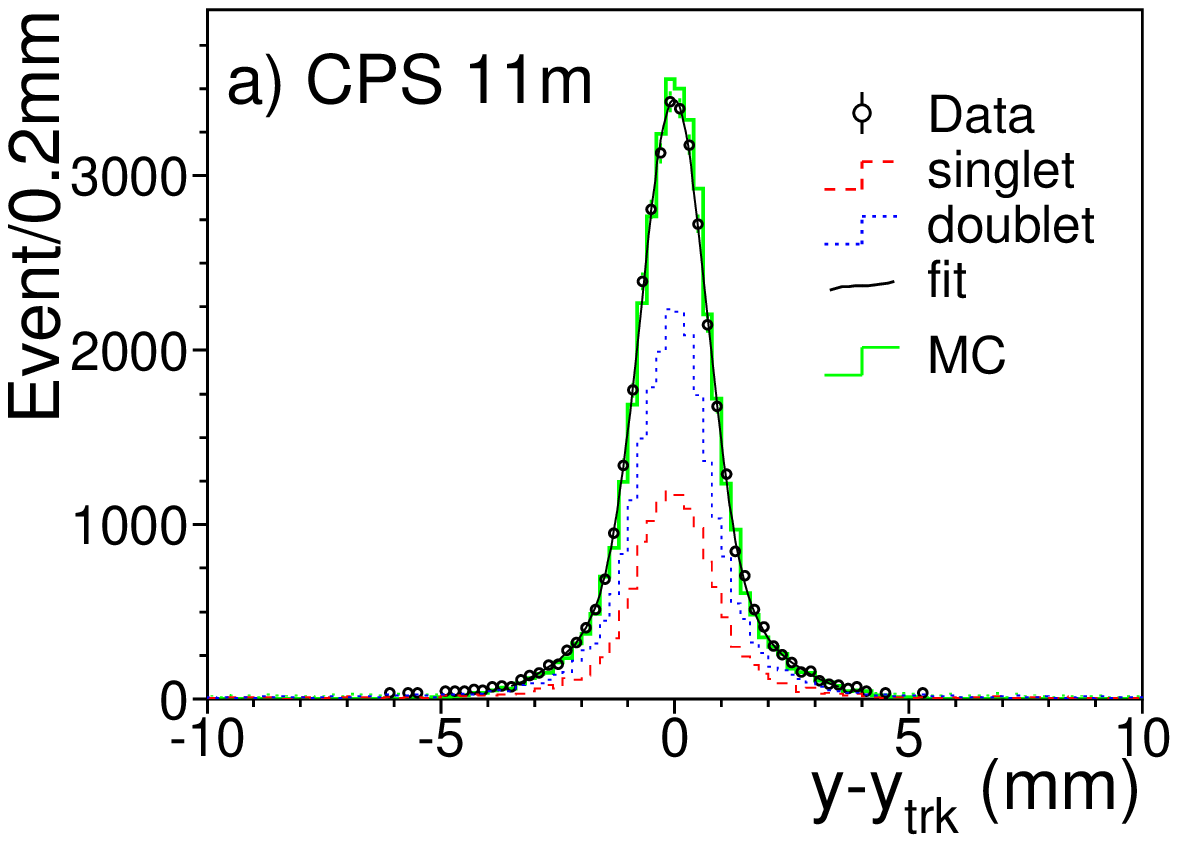}

  \vspace{-2.cm}
  \includegraphics[width=.75\linewidth] {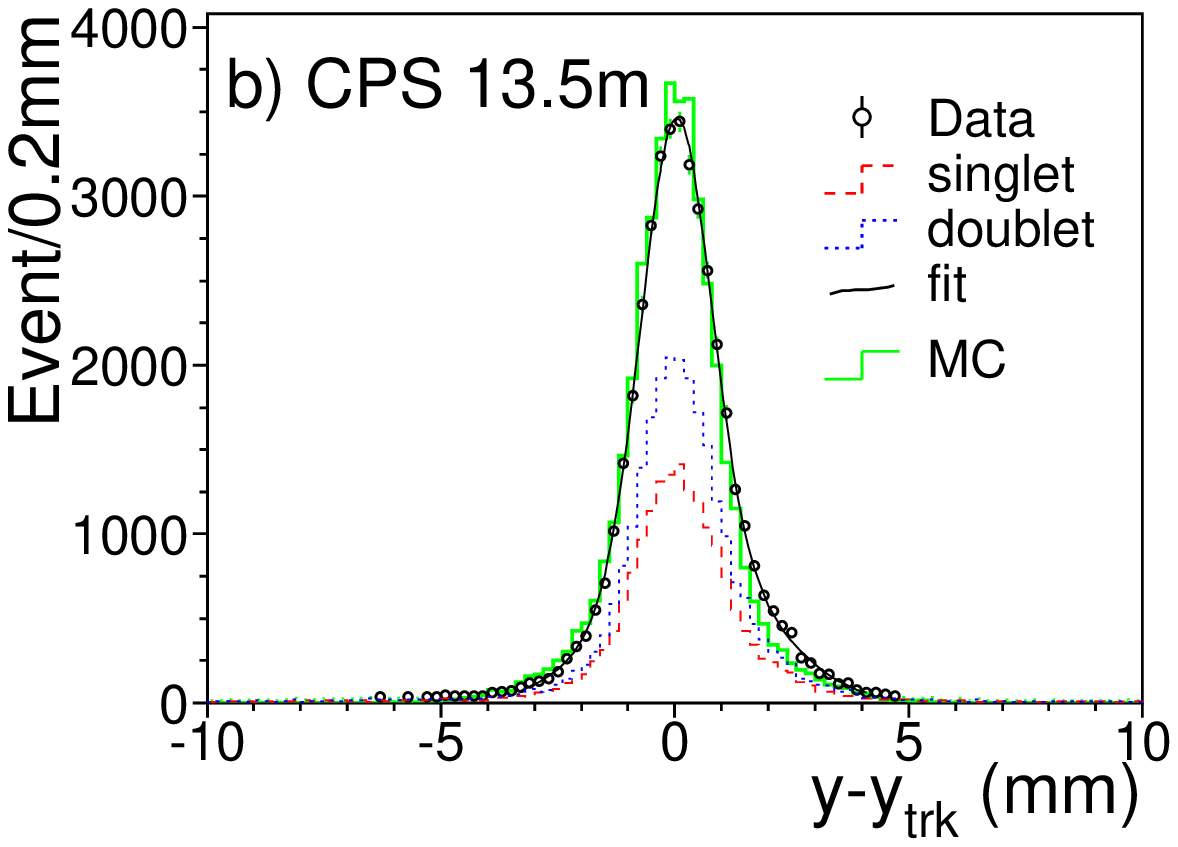}

  \vspace{-2.cm}
  \includegraphics[width=.75\linewidth] {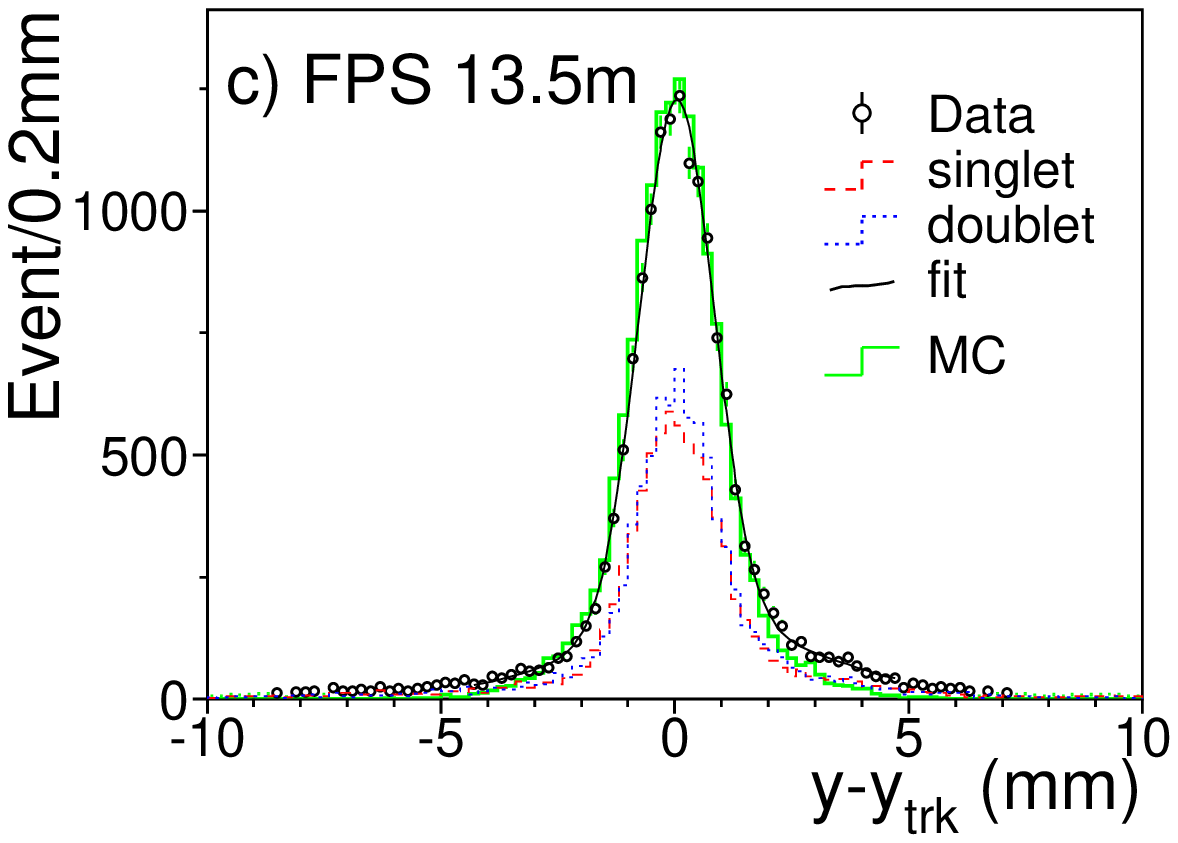}

  \vspace{-.5cm}
  \caption{Residuals of cluster positions relative to the PDT track
           for CPS with (a) 11 m and (b) 13.5 m clear fibers,
           and (c) FPS with 13.5 m clear fibers.
           The curves are double-Gaussian fits to data (circles).
           The contributions from singlets (dashed)
           and doublets (dotted lines) are also shown.
           GEANT simulations are shown by the histograms.
  \label{fig:resi} }
  \vspace{-.5cm}
\end{figure}                           %----------------------Fig.23

%====================================================
\subsection {GEANT simulation and spatial resolution }

The residuals for the positions of preshower clusters relative to
the PDT tracks are plotted in Fig.~\ref{fig:resi}
for the CPS with 11 m and 13.5 m clear fibers,
and for the FPS-$u$ (axial) layer with 13.5 m clear fibers.
The curves are fits of double-Gaussian shapes to the data.
The parameters from the fits are listed in Table~\ref{tab:resi}.
Results of separate fits to singlet and doublet contributions are 
also listed in Table~\ref{tab:resi}.

The widths of the narrower Gaussians are used to define
the detector resolution.
The wider Gaussians are employed to obtain good fits in the tails.
The lower light yield for the CPS with 13.5 m clear fibers
gives wider residuals than those with 11 m clear fibers.
The FPS has wider strip pitch and more singlet events,
which also result in wider distributions.
The small asymmetry seen in the distributions is caused
by the offset of trigger counters from the muon direction.
The dominant systematic uncertainties are contributed by 
multiple scattering and strip alignment.
The uncertainties in the survey of the strip positions in $z$
and the straightness of the strips have a large effect on 
the precision of the PDT track position.
The relative contributions of these factors are evaluated 
using a GEANT simulation.

The GEANT simulations calculate the ionization energy loss of the muon 
and use default energy cutoffs (1 \MeV) for bremsstrahlung photons
and electrons.
A detector hit is assigned at the mean of
the entrance and exit positions in an active volume.
To account for momentum dependence, the cosmic ray muon momentum
spectrum at sea level \cite{Allkofer,MASS} is generated and 
the filtering by the iron block is taken into account.
The incident muon angle is sampled according to that observed in
the data.
The angular deflection by multiple scattering is calculated 
using Moli\`ere theory \cite{Moliere1,Moliere2,Bethe,Scott}.
Detector resolutions are approximated by Gaussian smearings
of the hit positions for preshower detectors as well as for PDT hits.
The Gaussian widths representing detector resolutions are obtained
through a $\chi^2$ minimization of the simulated residuals 
relative to those in the data.

The PDT hits are assigned and processed independently
in the same way as the data, 
using the same event selection criteria and $\chi^2$ cutoffs.
The width of the Gaussian smearing applied to the PDT hits is
470 \um{}, which is 6\% lower than the nominal value \cite{PDT}.
The combined effect on the precision of the reference track positions 
from multiple scattering and the smearing of the PDT hits is estimated
to be 380 \um{} for the CPS and 390 \um{} for the FPS.
The muon momentum spectrum at sea-level as measured by 
different experiments has large uncertainties \cite{muon2}.
The uncertainty in the muon momentum spectrum contributes a 2\% 
systematic error to the resolution analysis.

The preshower light yield is simulated so as to resemble the data, with
the light yield spectrum generated as a function of the length of
scintillator traversed by the muon.
For doublet events, the spatial resolution is well described by a
Gaussian smearing at the hit position, and the width is determined
by adjusting it until the simulated residuals agree with the data.
The singlet events are assigned randomly over the geometric acceptance
at the strip center (Fig.~\ref{fig:eta_3}), which corresponds to 
a Gaussian distribution of about 400 \um{} in width.
The simulated residuals have an approximately
Gaussian shape with a width of 620 \um{}.
Gaussian smearing is also applied to singlets, and the spatial
resolution is obtained from a quadratic sum of the widths 
for the geometric acceptance (400 \um) and this Gaussian smearing.

The spatial resolutions estimated for ideal alignments are
listed in Table~\ref{tab:resolution}, where the combined resolutions
are given by the weighted sum of singlet and doublet resolutions.
The uncertainties in strip alignment and position in $z$
make large contributions to the residual distributions in the data.
The total is estimated to be 200 \um{}
predominantly from the 1 mm inaccuracy in the strip position in $z$.
To account for the geometric uncertainties,  
strip positions in the simulations are smeared by a Gaussian 
of 200 \um{} in width.
The residual distributions seen in the data are reproduced when
the geometric uncertainties are combined with the uncertainties in the
PDT track positions and the preshower resolutions listed in Table 3.
The best result is 510 \um{} for the CPS doublet events
using 11 m clear fibers.
The larger light yield gives a better signal-to-noise ratio 
and smaller cutoff effect from VLPC quantization for photoelectrons.
The systematic error is estimated to be 10\%, and is 
dominated by uncertainties in the alignment effects.

\begin{table}[t!]                 %-------------------------Table 3
  \begin{center}                  % was 00p031, 00p78, new 00p100
  \begin{tabular}{ll|ccc} 
  \hline
  \hline
   &&    Ideal geometry (\um) & Uncertainty $\sigma$=200 \um{} (\um)\\
  \hline
    CPS (11m)   & singlet &   $610\pm60$  & $580\pm60$  \\
    CPS (11m)   & doublet &   $550\pm60$  & $510\pm50$  \\
    CPS (11m)   & combined&   $570\pm60$  & $540\pm50$  \\
  \hline
    CPS (13.5m) & singlet &   $650\pm70$  & $620\pm60$  \\
    CPS (13.5m) & doublet &   $620\pm60$  & $580\pm60$  \\
    CPS (13.5m) & combined&   $630\pm60$  & $600\pm60$  \\
  \hline
    FPS (13.5m) & singlet &   $730\pm70$  & $700\pm70$  \\
    FPS (13.5m) & doublet &   $620\pm60$  & $590\pm60$  \\
    FPS (13.5m) & combined&   $680\pm70$  & $650\pm70$  \\
  \hline
  \hline
  \end{tabular}
  \caption{ Spatial resolutions estimated for ideal geometry 
            and the geometry having an uncertainty of 200 \um{}
            contributed by the inaccuracy of strip alignment 
            and position in $z$.
            The errors correspond to estimates of systematic
            uncertainties.
  \label{tab:resolution}  }
  \end{center}
  \vspace{.5cm}
\end{table}                       %--------------------------Table 3

%=============================================================Section 8
\section {Detection efficiency}

\begin{figure}[t!]                   %----------------------Fig.24
  \vspace{-1.cm}
  \centering
  \includegraphics[width=.6\linewidth] {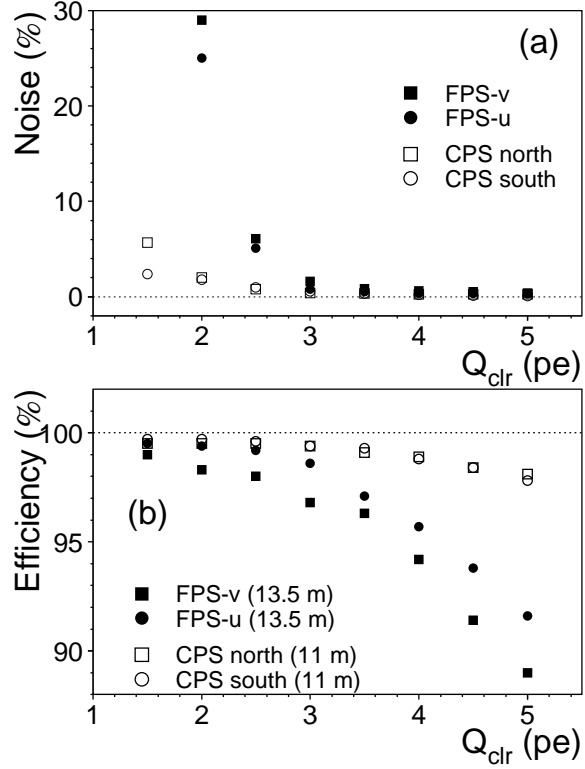} % {effi.eps}
  \vspace{-.5cm}
  \caption{(a) Event fractions of noise clusters and (b) detection 
           efficiencies as functions of photoelectron threshold
           for the CPS segments with 11 m clear fibers and
           the FPS layers with 13.5 m clear fibers.
  \label{fig:effi} }
  \vspace{.5cm}
\end{figure}

The detection efficiency of the preshower detectors
is studied relative to the PDT track position in a region
50 mm away from the ends of strips.
The CPS strips are searched for a hit at the expected position.
If a predicted hit is found, for example, on the CPS-north segment,
noise clusters are counted for CPS-south.
The FPS hits are examined in a similar way in
an overlapping region of the two layers.
When a hit is found in one layer, the other layer is searched 
for a hit.
A noise cluster is defined if the PDT track is far from the FPS module.
The noise contamination and the efficiency for detecting 
minimum ionizing particles are studied as functions of the 
threshold set on the yield for the strip cluster.
Plotted in Fig.~\ref{fig:effi}, as a example,
are the distributions obtained for the CPS segments
with 11 m clear fibers, and the FPS layers with 13.5 m clear fibers.
Several combinations of VLPC modules and clear fiber lengths
have been tested and the results for the CPS and the FPS are compatible.

The occurrence of noise clusters depends on the VLPC gain 
and the contribution from thermal electrons.
The distributions shown in Fig.~\ref{fig:effi}.a for the CPS segments
are obtained with low-gain chips ($\sim$9 ADC/pe) and 
on average below 5\% of events originated from thermal electrons.
The VLPCs used for the FPS module have moderate gains ($\sim$12 ADC/pe)
and higher fractions of events with thermal electrons ($>$5\%),
and the noise contamination increases significantly below a cluster
threshold of $Q_{clr}=$3 photoelectrons.

The detection efficiency for minimum ionizing particles
depends on the light yield. 
In Fig.~\ref{fig:effi}.b the CPS and the FPS results are
obtained with clear fibers of 11 m and 13.5 m in lengths, 
respectively.
The mean light yields for singlets are approximately 
11 and 8 photoelectrons, respectively.
At a cluster threshold of 3 photoelectrons, 
the detection efficiencies of both CPS and FPS have reached 99\%.
As the cluster threshold increases,
the loss of CPS efficiency is much less than that for FPS.
The effect of having a defective channel can be seen for 
the FPS-$v$ layer. 
The reduced efficiency is roughly proportional to the geometric 
fraction of singlets in a strip.

                         %----------------------Fig.24

%=============================================================Section 9
\section {Conclusion }

The \D0 preshower modules, tested with cosmic ray muons 
during three periods spread over six months time, 
have shown a long-term stability 
with consistent MIP light yields from three VLPC cassettes.
The mean light yields obtained with 11 m clear fibers is 12 
photoelectrons for singlet clusters and 14 photoelectrons 
for doublet clusters, and the strip-by-strip uniformity 
is better than 10\%.

The triangular cross section of the scintillator strips allows a 
convenient configuration for the reconstruction of track positions.
The spatial resolution is about 15\% of the strip pitch.
For the CPS doublets, with a mean light yield of 14 photoelectrons,
the spatial resolution is found to be $510\pm 50$ \um{}.
The MIP signals are well separated from electronic noise and
signals of thermal electrons.
At low light-yield thresholds, a 99\% detection efficiency 
can be achieved with only a few percent noise contamination.

%=======================================================Acknowledgements
\section*{Acknowledgements}
We wish to thank our colleagues on \D0 for their essential 
contributions.
We would like to express our gratitude to Prof.  T.\ Ferbel
for stimulating discussions and careful reading of the manuscript.
We acknowledge technical assistance provided by the 
\D0 department of the Fermilab Particle Physics Division.
This work was supported by the U.S. Department of Energy
and the National Science Foundation.

%=============================================================Reference
%\newpage

%=========================================================End contents

\begin{thebibliography}{}

  \bibitem{D0}
  S. Abachi et al., FERMILAB-PUB-96-357-E, October 1996.
 
  \bibitem{VLPC}
  D. Lincoln, Nucl. Phys. B (Proc. Suppl.) 78 (1999) 281,
  and references therein.

  \bibitem{sqstrip}
  M. Adams et al., Nucl. Instr. and Meth. A 366 (1995) 263.

  \bibitem{tristrip}
  M. Adams et al., Nucl. Instr. and Meth. A 378 (1996) 131.

  \bibitem{GEANT}
  GEANT Version 3.21 (October 1994);
  R. Brun et al., CERN DD/EE/84-1 (September 1987),
  CERN Program Library Long Writeup W5013 (October 1994).

  \bibitem{PDT}
  Yu.M. Antipov et al., Nucl. Instr. and Meth. A 297 (1990) 121.

  \bibitem{extrusion}
  A. Pla-Dalmau et al., FERMILAB-CONF-99-095, April 1999.

  \bibitem{Bicron}
  Bicron Corporation, 12345 Kinsman Road, Newbury, OH, 44065, USA.

  \bibitem{Kuraray}
  Kuraray Intl., 200 Park Ave, New York, NY 10166, USA.

  \bibitem{cryo}
  T.H. Gasteyer and P.D. Wheelwright,
  FERMILAB-PUB-97-255, July 1997.
 
  \bibitem{SVX}
  R. Yarema et al., FERMILAB-TM-1892, 
  October 1996, and references therein.

  \bibitem{sequencer}
  M. Utes, \D0 Engineering Note 3823.110-EN-504, March 1999.

  \bibitem{DART}
  Online system developed by the Online System Department,
  Fermilab Computing Division, FNAL PN509, July 1998.

  \bibitem{LFIT}
  M. Metcalf, LFIT, CERNLIB E250, November 1984.

  \bibitem{Allkofer}
  O.C. Allkofer et al., Phys. Lett. 36B (1971) 425.

  \bibitem{MASS}
  M.P. De Pascale et al., J. Geophys. Res. 98 (1993) 3501.

  \bibitem{Moliere1}
  G.Z. Moli\`ere, Z. Naturforsch. 2a (1947) 133.

  \bibitem{Moliere2}
  G.Z. Moli\`ere, Z. Naturforsch. 3a (1948) 78.

  \bibitem{Bethe}
  H.A. Bethe, Phys. Rev. 89 (1953) 1256.

  \bibitem{Scott}
  W.T. Scott, Rev. Mod. Phys. 35 (1963) 231.

  \bibitem{muon2}
  See, for example, discussions in 
  S. Cecchini and M. Sioli, DFUB-99-26(hep-ex/0002052),
  5th School on Non-accelerator Particle Astrophysics, 
  Trieste, Italy, 1998.

\end{thebibliography}
\end{document}